\documentclass[showpacs,preprintnumbers,amsmath,amssymb]{revtex4}
\usepackage[dvips]{graphicx}
\usepackage{dcolumn}

\begin{document}

\title{Strong Coupling Correction in Superfluid $^3$He in Aerogel}

\author{Kazushi Aoyama and Ryusuke Ikeda}

\affiliation{%
Department of Physics, Kyoto University, Kyoto 606-8502, Japan
}

\date{\today}

\begin{abstract}
Effects of impurity scatterings on the strong coupling (SC) contribution, stabilizing the ABM (axial) pairing state, to the quartic term of the Ginzburg-Landau (GL) free energy of superfluid $^3$He are theoretically studied to examine recent observations suggestive of an anomalously small SC effect in superfluid $^3$He in aerogels. To study the SC corrections, two approaches are used. One is based on a perturbation in the short-range repulsive interaction, and the other is a phenomenological approach used previously for the bulk liquid by Sauls and Serene [Phys.Rev.B {\bf 24}, 183 (1981)]. It is found that the impurity scattering favors the BW pairing state and shrinks the region of the ABM pairing state in the $T$-$P$ phase diagram. In the phenomenological approach, the resulting shrinkage of the ABM region is especially substantial and, if assuming an anisotropy over a large scale in aerogel, leads to justifying the phase diagrams determined experimentally. 
\end{abstract}


\maketitle

\section{Introduction}
The aerogel is a highly porous medium composed of SiO$_2$ strands with a porosity of about 98\% in the case of samples used in many experiments on superfluid $^3$He. A structural correlation length, $\xi_a$, of aerogel is identified with the typical distance between strands, i.e., $\xi_a\approx 30-100$nm. In the system of superfluid $^3$He in aerogel, the coherence length $\xi_0$ at zero temperature competes with $\xi_a$ in magnitude, and it will be necessary to consider two different regimes separately to understand roles of aerogel in this system. At shorter scales than $\xi_a$, the aerogel acts as an anisotropic scatterer inducing an uniaxially anisotropic scattering amplitude of $^3$He quasiparticles, while it acts as point scatterers for them at much larger scales than $\xi_a$. In any case, the aerogel plays roles of a pair-breaker for superfluid $^3$He with a $p$-wave pairing.

In liquid $^3$He in aerogel, it is now believed that (at least) two superfluid states occur. One is the A-like phase, and the other is the B-like phase \cite{Nazaretski, Gervais}. The latter has essentially the same properties as the bulk B phase and thus, is believed to be in the BW pairing state. In contrast, the A-like phase with an equal-spin pairing (ESP) has remarkably different features in the phase diagram from those of the bulk A phase which is in the ABM (axial) pairing state. Among them, remarkable ones are 1) a shift of polycritical point (PCP) to {\it lower} pressures, 2) the positive slope ($dT_{\rm AB}/dP > 0$), appearing {\it even at high pressures}, of the A-B transition curve $T_{\rm AB}(P)$ in contrast to the negative slope in the bulk case \cite{com1,Osheroff, Halperin}, and 3) an anomalously high A1-A2 transition field \cite{A1A2}. Based on the feature 1), a different pairing state from the ABM one was proposed \cite{Fomin}. However, it has been clarified recently that 1) can be understood, under the assumption that the A-like phase is in the ABM pairing state, as a consequence of an anisotropy existing over long distances in the aerogel sample \cite{AI2} or of an act of aerogel as a quenched disorder \cite{AI}. Although the features at lower pressures on the phase diagram in aerogel are well understood within these approaches \cite{AI2,AI}, the features 2) and 3) peculiar to the high pressure region remain to be explained. However, the point to be searched for seems to be clear: It is well understood \cite{BSA,RS} that the ABM pairing state in bulk is stabilized by a correction term to the quartic term in the Ginzburg-Landau (GL) free energy induced by the repulsive interaction in the normal state. This correction term, called as the strong-coupling (SC) correction, is enhanced with increasing the pressure $P$, because the effective repulsive interaction between quasiparticles will be stronger at higher $P$-values where the bulk A phase is present over a wider temperature range. On the other hand, in the liquid $^3$He in aerogel, the SC correction was estimated based on the observation 3) to be anomalously small, and such a small SC correction may be consistent with the positive slope in 2), because the positive slope may be realized if the temperaure width of the A-like phase at the high pressure end of the liquid phase is narrow enough \cite{Osheroff,Halperin}. So far, the SC correction in the impure superfluid $^3$He was incorporated in theoretical calculations using a simplified treatment in which impurity effects have been incorporated merely through a relaxation rate of quasiparticles \cite{Georgia,AI}. However, the observations 2) and 3) seem to imply that this relaxation time approximation is insufficient if the A-like phase is in the ABM pairing state.

 In this paper, the impurity effects on the SC correction are studied in detail and carefully on the basis of two analytical methods, a purely diagrammatic approach \cite{rep} treating the four-point vertex (FPV) of the normal quasiparticles perturbatively in the strength of a bare repulsive interaction, and a phenomenological approach \cite{SS} in which the FPV is parametrized in terms of Landau parameters estimated from properties in the normal liquid $^3$He. We find that, in both of the two approaches, the SC correction is reduced by incorporating an impurity-induced new process in the SC correction, and the resulting shrinkage of the region of the ABM pairing state in the impure case is much more remarkable in the phenomenological approach, suggesting that this approach \cite{SS} is more suitable to a description of real liquid $^3$He.

In sec.II, the details of our theoretical analysis are explained, and the resulting general formulation is applied in the ensuing two subsections to examine the phase diagram of weakly disordered $^3$He according to the two approaches. In sec.III, the obtained results are used to discuss the experimentally determined phase diagrams of the liquid $^3$He in aerogel by incorporating possible anisotropies in the impurity scattering provided by the aerogel structures. Concluding remarks are given in the final section.

\section{Microscopic Calculation of Strong Coupling Correction}
The superfluid $^3$He is in a spin triplet p-wave pairing state with a gap function of the form 
\begin{eqnarray}
\Delta_{\alpha\beta}({\hat {\bf p}}) = \vec{\Delta}({\hat {\bf p}}) \cdot (i \, \vec{\sigma} \, \sigma_2)_{\alpha\beta} = A_{\mu,j} {\hat p}_j \, (i \, \sigma_\mu \, \sigma_2)_{\alpha\beta}, 
\end{eqnarray}
where $\vec{\sigma} = (\sigma_1, \, \sigma_2, \, \sigma_3)$ are Pauli matrices, ${\hat {\bf p}}$ is the unit vector ${\bf p}/p_F$ parallel to the momentum ${\bf p}$, and $p_F$ is the Fermi momentum. Since the aerogel structure acts as a nonmagnetic impurity in superfluid $^3$He, we start from the BCS Hamiltonian with an attractive $p$-wave pairing interaction and an impurity scattering term, i.e., ${\hat H}_{\rm BCS}={\hat H}_{p}+{\hat H}_{\rm imp}$. Here, 
\begin{eqnarray}
{\hat H}_{p} - \mu {\hat N} &=& \sum_{{\bf p}, \alpha} \biggl[ \frac{p^2}{2m}-\mu \biggr] {\hat a}^\dagger_{{\bf p},\alpha} \, {\hat a}_{{\bf p},\alpha} - 3 g_1 \sum_{\bf q} {\hat O}_{j, \mu}^\dagger({\bf q}) \, {\hat O}_{j,\mu}({\bf q}), \nonumber \\
H_{\rm imp} &=& \sum_{\alpha}\int_{\bf r} {\hat \Psi}^{\dagger}_{\alpha}({\bf r})u({\bf r}) {\hat \Psi}_{\alpha}({\bf r}), \nonumber \\
{\hat O}_{j,\mu}({\bf q}) &=& \sum_{\bf p} \frac{p_j}{p_F} \, {\hat a}_{-{\bf p}+{\bf q}/2, \alpha} ({\rm i} \, \sigma_\mu \, \sigma_2)_{\alpha \beta} \, {\hat a}_{{\bf p}+{\bf q}/2, \beta},
\end{eqnarray}
where ${\hat a}_{{\bf p},\alpha}$ is the Fourier transform of the field operator $\Psi_\alpha({\bf r})$ of quasiparticles, $u({\bf r})$ is an impurity scattering potential with a Gaussian ensemble defined by ${\overline {u({\bf r})}}=0$ and ${\overline {u({\bf r}_1) u({\bf r}_2)}}= (2 \pi \tau N(0))^{-1} \delta({\bf r}_1 - {\bf r}_2)$, the overbar implies the random average, $\tau$ is the life time of a quasiparticle, and $N(0)$ is the density of states per spin on the Fermi surface. The $p$-wave pairing channel has been assumed to be dominant in the interaction Hamiltonian 
\begin{equation}
H_{QP}=\int_{\bf r} \int_{{\bf r}'} \, [ \, V^{(s)}({\bf r}-{\bf r}') {\hat n}({\bf r}) {\hat n}({\bf r}') + V^{(a)}({\bf r}-{\bf r}') {\hat s}_\mu({\bf r}) {\hat s}_\mu({\bf r}') \, ],
\end{equation}
between quasiparticles conserving the total spin, where ${\hat n}$ and ${\hat s}_\mu$ are the density and spin-density operators of quasiparticles, 
respectively.

Throughout this paper, we work in the mean field approximation where the roles \cite{AI} of the impurity scattering acting as a quenched disorder on the superfluid order parameter $A_{\mu,j}({\bf r})$ are neglected, and thus, spatial variations of $A_{\mu,j}$ may be neglected in considering the equilibrium properties. Further, in examining impurity effects on the SC correction, a possible anisotropy in the scattering events will be neglected since, as well as the anisotropy, the SC correction itself is a small contribution to the condensation energy. 
Inclusion of anisotropy of scattering events will be postponed to sec.III in which results comparable with an experimental phase diagram will be shown. Then, we only have to examine the coefficients of GL free energy functional 
\begin{eqnarray}\label{eq:bulk}
{\cal H}_{\rm GL}&=&\int_{\bf r}\Big(\alpha A_{\mu,i}^\ast A_{\mu,i}+\beta_1 |A_{\mu,i}^{\ast}A_{\mu,i}|^2+\beta_2 (A_{\mu,i}^{\ast}A_{\mu,i})^2+\beta_3 A_{\mu,i}^{\ast}A_{\nu,i}^{\ast}A_{\mu,j}A_{\nu,j}\nonumber\\
&&+\beta_4 A_{\mu,i}^{\ast}A_{\nu,i}A_{\nu,j}^{\ast}A_{\mu,j}+\beta_5 A_{\mu,i}^{\ast}A_{\mu,i}A_{\nu,j}A_{\mu,j}^{\ast} \Big)
\end{eqnarray}
microscopically. When the Born approximation for $u({\bf r})$ is used, and the $s$-wave component of the scattering amplitude is assumed to be dominant, the coefficient $\alpha$ in eq.(4) may be well approximated by the familiar result 
\begin{eqnarray}
\alpha&=&\frac{1}{3}N(0)\big[\ln\frac{T}{T_{c0}}+\psi\big(\frac{1}{2}+\frac{1}{4\pi\tau T}\big)-\psi\big(\frac{1}{2}\big)\big], \nonumber 
\end{eqnarray}
where $T_{c0}$ is the superfluid transition temperature of the bulk liquid in the mean field approximation. 
Note that, due to the ${\bf p}$-dependence carried by the the pair-field vertex, the impurity-induced vertex sketched in Fig.1(b) can be neglected, where a solid line is the quasiparticle Green's function and a dashed line with a cross symbol represents the impurity scattering. In the weak coupling approximation, other coefficients in eq.(4), obtained consistently with the above $\alpha$, are 
given by
\begin{eqnarray}
\beta_3&=&\beta_2=\beta_4=-\beta_5 =-2\beta_1=2\beta_{\rm wc}=-\frac{\beta_0(T)}{7 \zeta(3)}\psi^{(2)}\big(\frac{1}{2}+\frac{1}{4\pi\tau T}\big), \nonumber \\ 
\beta_0(T) &=& \frac{7 \zeta(3) N(0)}{240 \pi^2 T^2}, 
\end{eqnarray}
where $\psi^{(n)}(z)$ is diagamma function. 
\begin{figure}[t]
\includegraphics[scale=0.4]{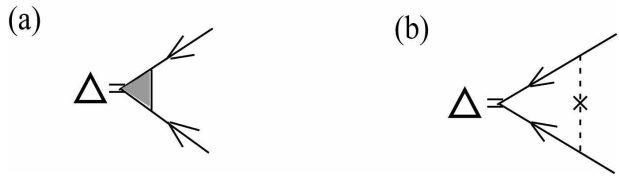}
\caption{(a) Vertex (shaded part) between two quasiparticles and the pair-field $\Delta$. (b) Example of the vertex with an impurity-induced correction which vanishes in the $p$-wave pairing when the scattering amplitude $\tau^{-1}$ is independent of the momenta. The dashed line with a cross carries $(2 \pi N(0) \tau)^{-1}$. \label{fig:vertex}}
\end{figure}

The most stable pairing state is determined by comparing the condensation energy $f^{\rm cond}=-\alpha^2/(4\beta_{\rm N})$ between different pairing states with one another, where the index ${\rm N}$ indicates each pairing state. Since $\alpha$ is common to different pairing states, the stability depends only on $\beta_{\rm N}$. We investigate the stability of the ABM pairing state relative to the BW one, that is, the relative difference between $\beta_{\rm ABM}\equiv \beta_{245}$ and $\beta_{\rm BW}\equiv \beta_{12}+\beta_{345}/3$, where $\beta_{12}=\beta_1+\beta_2$, and $\beta_{ijk}=\beta_i+\beta_j+\beta_k$.

\begin{figure}[t]
\includegraphics[scale=0.47]{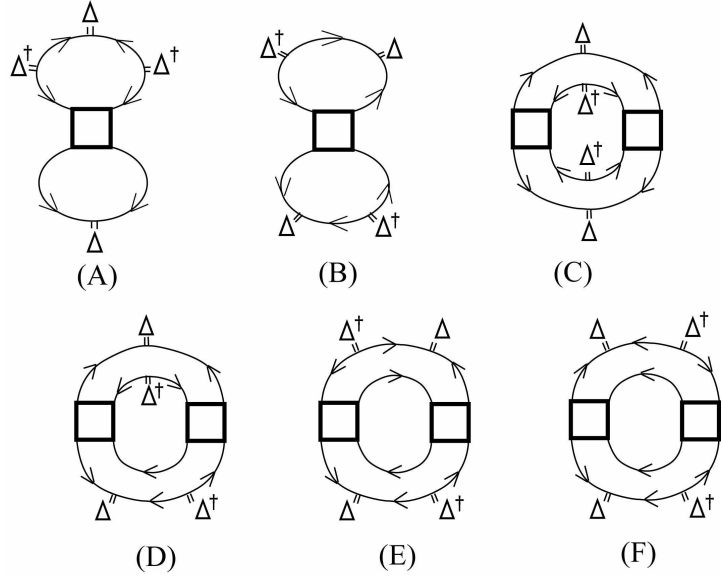}
\caption{Diagrams describing SC corrections in clean limit. Here, any frequency dependence in the FPV (the open square) is neglected. Diagrams (A) and (B) can be neglected for the reasons explained in the text. \label{fig:strong}}
\end{figure}

First, we consider the SC correction to $\beta_j$ in clean limit based on the general framework given by Rainer and Serene (RS) \cite{RS}. We will not use the conventional spin fluctuation model \cite{BSA}, because this model does not explain a property of the bulk B-phase: As can be seen in Table 5.1 of Ref.\cite{VW}, the pressure dependence of $\beta_{\rm BW}$ in the spin fluctuation model is opposite to the result \cite{SS} based on the RS's method and on the use of experimentally determined Landau parameters. The latter approach \cite{SS} seems to have explained the pressure dependence of the specific heat jump at the normal to B-phase transition \cite{cutoff}. According to Ref.\cite{RS}, the SC contribution to $\beta_j$, $\delta\beta_j$, can be obtained up to the lowest order in $T/E_F$ by examining diagrams raised in Fig.\ref{fig:strong}, where an open square is the quasiparticle 4-point vetex (FPV) function in the $T=0$ limit, $\Gamma^{(4)}_{\alpha\beta;\gamma\delta}({\bf p}_1,{\bf p}_2;{\bf p}_3,{\bf p}_4)$. Here, the renormalzed interaction Hamiltonian was assumed to take the form \cite{AGD} 
\begin{equation}
H_{\rm int}^{({\rm ren})} = \frac{1}{4} \sum_{\alpha,\beta,\gamma,\delta} \int_{\bf r_1}\int_{\bf r_2}\int_{\bf r_3}\int_{\bf r_4} \Psi^{\dagger}_{\alpha}({\bf r}_1) \Psi^{\dagger}_{\beta}({\bf r}_2) \Gamma^{(4)}_{\alpha \beta; \gamma \delta}({\bf r}_1,{\bf r}_2; {\bf r}_4, {\bf r}_3)  \Psi_{\delta}({\bf r}_4) \Psi_{\gamma}({\bf r}_3). 
\end{equation}
Among the diagrams in Fig.\ref{fig:strong}, the former two diagrams can be neglected :  Fig.2(A) belongs to the weak coupling diagram and expresses an interaction-induced renormalization of a pair-field vertex appearing commonly to all superfluid pairing states. This vertex correction has been examined in eq.(3.7) of Ref.\cite{RS} and estimated to be negligibly small compared with the SC corrections. Further, Fig.2(B) vanishes after carrying out frequency summations. Then, we have only to calculate the remaining four diagrams. Their contributions to the GL free energy functional are expressed as 
\begin{eqnarray}
{\rm Fig.(C)}&=&-\frac{1}{8}T^3\sum_{n_1,n_2,n_3}\int_{{\bf p}_1}\int_{{\bf p}_2}\int_{{\bf p}_3}G^{(0)}_{\varepsilon_1}({\bf p}_1)G^{(0)}_{-\varepsilon_1}(-{\bf p}_1)G^{(0)}_{\varepsilon_2}({\bf p}_2)G^{(0)}_{-\varepsilon_2}(-{\bf p}_2)G^{(0)}_{\varepsilon_3}({\bf p}_3)G^{(0)}_{-\varepsilon_3}(-{\bf p}_3)G^{(0)}_{\varepsilon_4}({\bf p}_4)G^{(0)}_{-\varepsilon_4}(-{\bf p}_4) \nonumber\\
&&\times \Gamma^{(4)}_{\alpha_1,\alpha_2;\alpha_3,\alpha_4}({\bf p}_1,{\bf p}_2;{\bf p}_3,{\bf p}_4)\Gamma^{(4)}_{\beta_1,\beta_2;\beta_3,\beta_4}(-{\bf p}_1,-{\bf p}_2;-{\bf p}_3,-{\bf p}_4)\Delta_{\alpha_1 \beta_1}({\hat p_1}) \Delta_{\alpha_2 \beta_2}({\hat p_2}) \Delta^{\dagger}_{\beta_3 \alpha_3}({\hat p_3}) \Delta^{\dagger}_{\beta4 \alpha_4}({\hat p_4}), \nonumber\\
{\rm Fig.(D)}&=&T^3\sum_{n_1,n_2,n_3}\int_{{\bf p}_1}\int_{{\bf p}_2}\int_{{\bf p}_3}G^{(0)}_{\varepsilon_1}({\bf p}_1)G^{(0)}_{-\varepsilon_1}(-{\bf p}_1) [ G^{(0)}_{\varepsilon_2}({\bf p}_2) ]^2 G^{(0)}_{-\varepsilon_2}(-{\bf p}_2)G^{(0)}_{\varepsilon_3}({\bf p}_3)G^{(0)}_{-\varepsilon_3}(-{\bf p}_3)G^{(0)}_{\varepsilon_4}({\bf p}_4) \nonumber\\
&&\times \Gamma^{(4)}_{\alpha_1,\alpha_2;\alpha_3,\alpha_4}({\bf p}_1,{\bf p}_2;{\bf p}_3,{\bf p}_4)\Gamma^{(4)}_{\beta_1,\alpha_4;\beta_3,\beta_4}(-{\bf p}_1,{\bf p}_4;-{\bf p}_3,{\bf p}_2)\Delta_{\alpha_1 \beta_1}({\hat p_1}) \Delta_{\alpha_2 \gamma}({\hat p_2})\Delta^{\dagger}_{\gamma \beta_4}({\hat p_2}) \Delta^{\dagger}_{\beta_3 \alpha_3}({\hat p_3}), \nonumber\\
{\rm Fig.(E)}&=&-\frac{1}{2}T^3\sum_{n_1,n_2,n_3}\int_{{\bf p}_1}\int_{{\bf p}_2}\int_{{\bf p}_3} [ G^{(0)}_{\varepsilon_1}({\bf p}_1) ]^2 G^{(0)}_{-\varepsilon_1}(-{\bf p}_1) [ G^{(0)}_{\varepsilon_3}({\bf p}_3) ]^2 G^{(0)}_{-\varepsilon_3}(-{\bf p}_3)G^{(0)}_{\varepsilon_2}({\bf p}_2)G^{(0)}_{\varepsilon_4}({\bf p}_4) \nonumber\\
&&\times \Gamma^{(4)}_{\alpha_1,\alpha_2;\alpha_3,\alpha_4}({\bf p}_1,{\bf p}_2;{\bf p}_4,{\bf p}_3)\Gamma^{(4)}_{\beta_1,\alpha_3;\alpha_2,\beta_4}({\bf p}_3,{\bf p}_4;{\bf p}_2,{\bf p}_1)\Delta_{\alpha_1 \gamma}({\hat p_1}) \Delta^{\dagger}_{\gamma \beta_4}({\hat p_1})\Delta_{\beta_1 \delta}({\hat p_3})\Delta^{\dagger}_{\delta \alpha_4}({\hat p_3}) , \nonumber\\
{\rm Fig.(F)}&=&-\frac{1}{4}T^3\sum_{n_1,n_2,n_3}\int_{{\bf p}_1}\int_{{\bf p}_2}\int_{{\bf p}_3} [ G^{(0)}_{\varepsilon_1}({\bf p}_1) ]^2 G^{(0)}_{-\varepsilon_1}(-{\bf p}_1) [ G^{(0)}_{\varepsilon_2}({\bf p}_2) ]^2 G^{(0)}_{-\varepsilon_2}(-{\bf p}_2)G^{(0)}_{\varepsilon_3}({\bf p}_3)G^{(0)}_{\varepsilon_4}({\bf p}_4) \nonumber\\
&&\times \Gamma^{(4)}_{\alpha_1,\alpha_2;\alpha_3,\alpha_4}({\bf p}_1,{\bf p}_2;{\bf p}_3,{\bf p}_4)\Gamma^{(4)}_{\alpha_3,\alpha_4;\beta_1,\beta_2}({\bf p}_3,{\bf p}_4;{\bf p}_1,{\bf p}_2)\Delta_{\alpha_1 \gamma}({\hat p_1}) \Delta^{\dagger}_{\gamma \beta_1}({\hat p_1})\Delta_{\alpha_2 \delta}({\hat p_3}) \Delta^{\dagger}_{\delta \beta_2}({\hat p_3}), \nonumber
\end{eqnarray}
where $G^{(0)}_{\varepsilon_n}({\bf p})$ is the quasiparticle Green's function in the normsal state in clean limit, ${\bf p}_4={\bf p}_1+{\bf p}_2-{\bf p}_3$, and the integral $\int_{\bf p}$ means $\int d^3p/(2\pi)^3$.

Now, let us turn to the case with effects of impurity scattering. In each process accompanying impurity scatterings, we will keep only the lowest order term in the impurity scattering: In fact, as a consequence of the process dipicted in Fig.1(b), the non $s$-wave paired superfluid is destroyed when $\tau T_{c0}$ is of order unity so that we need to focus on the case of a weak impurity scattering in which $\tau T_{c0} > 1$. Further, the assumption of a weak impurity scattering is reasonable in the sense that the aerogel is a porous medium with so dilute scattering centers. First, for single particle properties such as the selfenergy term of the quasiparticle Green's function, we use the Born approximation by neglecting multiple scattering processes. Then, this self energy correction can be incorporated in the expressions of the GL terms by replacing the Matsubara frequency $\varepsilon_n$ in $G^{(0)}_{\varepsilon_n}({\bf p}) \equiv G^{(0)}(\varepsilon_n)$ 
by ${\tilde \varepsilon}_n \equiv \varepsilon_n( \, 1 + 1/(2 \tau |\varepsilon_n|) \, )$. Hereafter, a quasiparticle Green's function will be expressed by the resulting one $G_{\varepsilon_n} \equiv G^{(0)}({\tilde \varepsilon}_n)$. If other impurity effects are not considered, the resulting expressions of $\delta \beta_j$ in this relaxation time approximation are equivalent to those used in previous works \cite{Georgia,AI}. 

Here, let us first discuss mean field phase diagrams following from this relaxation time approximation. When determining an AB transition curve $T_{AB}(P)$, the next 6th order term has to be taken into account in the GL free energy in the weak coupling approximation where different coefficients $\beta_j$s have the same temperature dependence proportional to $T^{-2}$ as one another. However, the SC correction to $\beta_j$ brings additional dependences on $T$ and $P$, and, at least close to $T_c(P)$, a $T_{AB}(P)$-curve may be determined by eq.(4) with the SC correction but with no 6th order term. As in previous works \cite{BSA,KN} determining the bulk $T_{AB}$ line, it is natural to expect both the 6th order term and the SC correction to have to be incorporated in calculations. However, we have found that, even for the bulk $^3$He with $T_{AB}(P)$ far apart from $T_{c0}(P)$, reasonable $T_{AB}(P)$ curves can be obtained from eq.(4) with no 6th order term in the two approaches to be explained later. For this reason, we determine hereafter the phase diagram according to the GL free energy, eq.(4), truncated at the 4th order in $A_{\mu,i}$. 

In Fig.3, dependences of the width of the A-like region on $(\tau T_{c0})^{-1}$ resulting from this relaxation time approximation are shown. There, the $T_{AB}(P)$ curve is found to rather decrease with increasing the disorder strength $(\tau T_{c0})^{-1}$. It implies that the SC correction in this relaxation time approximation is rather enhanced with increasing disorder: Based on eq.(4), the $T_{AB}$ line is determined by the relation $\beta_{\rm ABM} = \beta_{\rm BW}$, with the SC corrections to $\beta_j$ included, which is independent of $T_c(P)$ determined from $\alpha$ in eq.(4). Hence, the impurity-induced reduction of $T_{\rm AB}$ seen in the $\tau$-dependence of Fig.3 implies an extension of the ABM state region if the familiar impurity-induced reduction of $T_c(P)$ is absent. Once considering correspondences with experimental facts suggesting a remarkable impurity-induced reduction of the SC parameter \cite{A1A2} (see sec.I), we feel that an impurity-induced reduction of the SC parameter will be present and needed to describe superfluid $^3$He in aerogel properly. 

\begin{figure}[h]
\includegraphics[scale=0.6]{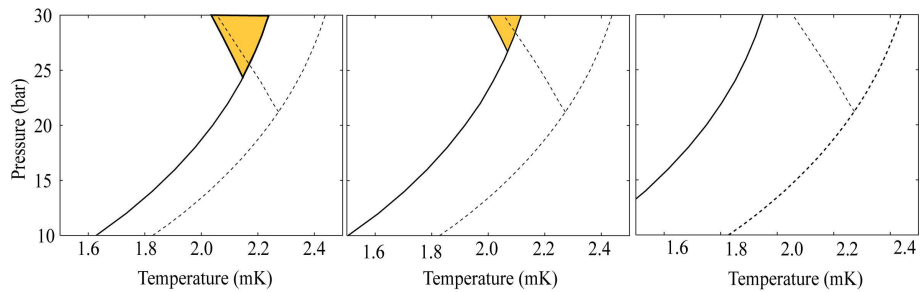}
\caption{Mean field phase diagrams in the case with {\it no anisotropy} and in the relaxation time approximation obtained in terms of $(2 \pi \tau)^{-1}=0.08$ (left), $0.13$, and $0.19$ (right) (mK). The shaded regions denote those of the mean field A-like phase appearing without effects of anisotropies and the quenched disorder. The AB transition curve affected by the impurity scatterings (solid curve) is shifted to {\it lower} temperatures as the impurity concentration $\sim (T_{c0} \tau)^{-1}$ is increased. For comparison, the bulk AB transition line determined experimentally is indicated by a dashed curve in each case. \label{}}
\end{figure}

\begin{figure}[t]
\includegraphics[scale=0.39]{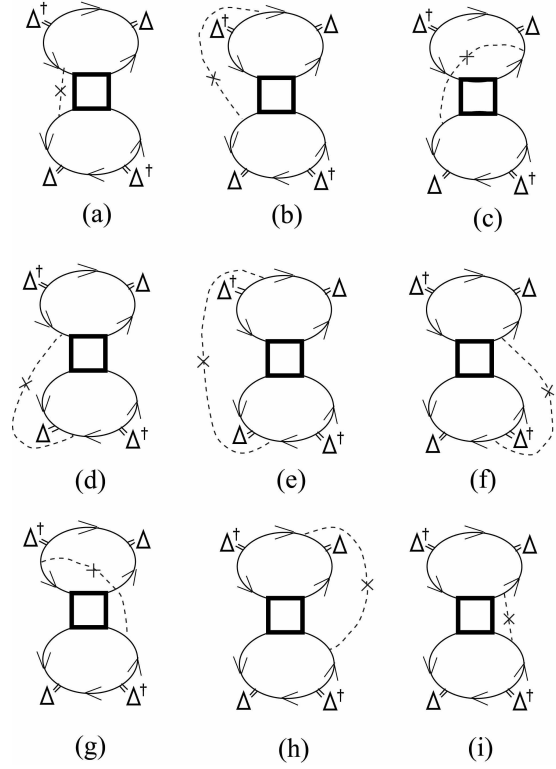}
\caption{Diagrams describing the impurity-induced SC corrections which do not appear in the relaxation time approximation. Note that the diagrams (a), (b), and (d) are identical with (i), (f) and (h), respectively. \label{fig:strongimp}}
\end{figure}

For the reason mentioned above, we investigate hereafter impurity-induced corrections to the two-particle processes, which include impurity-induced vertex corrections and will be denoted as $\delta {\tilde \beta}_j$ later. Among such vertex correction terms, the lowest order terms in $(\tau T_{c0})^{-1}$ contributing to $\delta {\tilde \beta}_j$ are obtained by replacing one of two FPVs in Fig.2C - 2F with a bare impurity vertex or adding an impurity line to Fig.2A and 2B. The resulting nonvanishing diagrams are those in Fig.\ref{fig:strongimp}. All diagrams in Fig.\ref{fig:strongimp} are obtained by adding a single impurity line to Fig.2B. On the other hand, the corresponding ones occurring from Fig.2A, which are sketched in Fig.5, vanish or can be neglected: Figure 5(a) can be absorbed into Fig.2A itself which, as already mentioned, is negligible. The next Fig.5(b) is absorbed into an additional weak-coupling quartic term including a single impurity line which was denoted as $-\varepsilon_{\rm imp} \beta_0$ in Ref.\cite{AI}. As noted there \cite{AI}, however, this $\varepsilon_{\rm imp}$-term does not contribute at all to the relative stability between different pairing states and hence, can be neglected in determining $T_{AB}(P)$ according to eq.(4). Further, Fig.5(c) is found to vanish at least up to the lowest order in $T/E_F$. This is verified by, following Ref.\cite{RS}, replacing the Green's function and their products appearing outside the FPVs in the way 
\begin{eqnarray}\label{eq:GF}
G_{\varepsilon}({\bf p}) &\to& -i\pi{\rm sgn} \varepsilon\delta(\xi_{\bf p}), \nonumber\\
G_{\varepsilon}({\bf p})G_{-\varepsilon}(-{\bf p}) &\to& \pi\frac{{\rm sgn}\varepsilon}{\tilde{\varepsilon}}\delta(\xi_{\bf p}), \nonumber\\
G^2_{\varepsilon}({\bf p})G_{-\varepsilon}(-{\bf p}) &\to& -i\frac{\pi}{2}\frac{{\rm sgn}\varepsilon}{\tilde{\varepsilon}^2}\delta(\xi_{\bf p}) 
\end{eqnarray}
and performing necessary momentum integrals and frequency summations. Further, according to the "rule" of the $T/E_F$-expansion formulated in Ref.\cite{RS}, the corresponding diagrams arising from Fig.2C-F and with a single impurity 
line are of higher order in $T/E_F$ compared with those of Fig.4 because an impurity line plays roles of a FPV with no finite frequencies carried. In this way, for our purposes of examining the SC corrections to $\beta_j$ up to the lowest order in $T/E_F$, we only have to focus on the diagrams in Fig.4 as the impurity effects on the SC correction which were not considered in the relaxation time approximation. 
Their contributions to the quartic term of the GL free energy are given by 

\begin{eqnarray}
{\rm Fig(a)}&=&\frac{1}{2}T\sum_{n_1}T\sum_{n_2}\int_{{\bf p}_1}\int_{{\bf p}_2}\int_{{\bf p}_3}G^2_{\varepsilon_1}({\bf p}_1)G_{-\varepsilon_1}(-{\bf p}_1)G_{\varepsilon_1}({\bf p}_3)G^2_{\varepsilon_2}({\bf p}_4)G_{-\varepsilon_2}(-{\bf p}_4)G_{\varepsilon_2}({\bf p}_2) \nonumber\\
&&\times |u({\bf p}_1-{\bf p}_3)|^2\Gamma^{(4)}_{\alpha_1,\alpha_2;\alpha_3,\alpha_4}({\bf p}_1,\varepsilon_1,{\bf p}_2,\varepsilon_2;{\bf p}_3,\varepsilon_1,{\bf p}_4,\varepsilon_2)\Delta_{\alpha_1 \beta}({\hat p_1})\Delta^{\dagger}_{\beta \alpha_3}({\hat p_1}) \Delta_{\alpha_2 \gamma}({\hat p_4}) \Delta^{\dagger}_{\gamma \alpha_4}({\hat p_4}), \nonumber\\
{\rm Fig(b)}&=&\frac{1}{2}T\sum_{n_1}T\sum_{n_2}\int_{{\bf p}_1}\int_{{\bf p}_2}\int_{{\bf p}_3}G_{\varepsilon_1}({\bf p}_1)G_{-\varepsilon_1}(-{\bf p}_1)G_{\varepsilon_1}({\bf p}_3)G_{-\varepsilon_1}(-{\bf p}_3)G^2_{\varepsilon_2}({\bf p}_4)G_{-\varepsilon_2}(-{\bf p}_4)G_{\varepsilon_2}({\bf p}_2) \nonumber\\
&&\times |u({\bf p}_1-{\bf p}_3)|^2\Gamma^{(4)}_{\alpha_1,\alpha_2;\alpha_3,\alpha_4}({\bf p}_1,\varepsilon_1,{\bf p}_2,\varepsilon_2;{\bf p}_3,\varepsilon_1,{\bf p}_4,\varepsilon_2) \Delta_{\alpha_1 \beta}({\hat p_1})\Delta^{\dagger}_{\beta \alpha_3}({\hat p_3}) \Delta_{\alpha_2 \gamma}({\hat p_4}) \Delta^{\dagger}_{\gamma \alpha_4}({\hat p_4}), \nonumber\\
{\rm Fig(c)}&=&\frac{1}{2}T\sum_{n_1}T\sum_{n_2}\int_{{\bf p}_1}\int_{{\bf p}_2}\int_{{\bf p}_3}G^2_{\varepsilon_1}({\bf p}_3)G_{-\varepsilon_1}(-{\bf p}_3)G_{\varepsilon_1}({\bf p}_1)G^2_{\varepsilon_2}({\bf p}_4)G_{-\varepsilon_2}(-{\bf p}_4)G_{\varepsilon_2}({\bf p}_2) \nonumber\\
&&\times |u({\bf p}_1-{\bf p}_3)|^2\Gamma^{(4)}_{\alpha_1,\alpha_2;\alpha_3,\alpha_4}({\bf p}_1,\varepsilon_1,{\bf p}_2,\varepsilon_2;{\bf p}_3,\varepsilon_1,{\bf p}_4,\varepsilon_2) \Delta_{\alpha_1 \beta}({\hat p_3})\Delta^{\dagger}_{\beta \alpha_3}({\hat p_3}) \Delta_{\alpha_2 \gamma}({\hat p_4}) \Delta^{\dagger}_{\gamma \alpha_4}({\hat p_4}), \nonumber\\
{\rm Fig(d)}&=&\frac{1}{2}T\sum_{n_1}T\sum_{n_2}\int_{{\bf p}_1}\int_{{\bf p}_2}\int_{{\bf p}_3}G^2_{\varepsilon_1}({\bf p}_1)G_{-\varepsilon_1}(-{\bf p}_1)G_{\varepsilon_1}({\bf p}_3)G_{\varepsilon_2}({\bf p}_4)G_{-\varepsilon_2}(-{\bf p}_4)G_{\varepsilon_2}({\bf p}_2)G_{-\varepsilon_2}(-{\bf p}_2) \nonumber\\
&&\times |u({\bf p}_1-{\bf p}_3)|^2\Gamma^{(4)}_{\alpha_1,\alpha_2;\alpha_3,\alpha_4}({\bf p}_1,\varepsilon_1,{\bf p}_2,\varepsilon_2;{\bf p}_3,\varepsilon_1,{\bf p}_4,\varepsilon_2) \Delta_{\alpha_1 \beta}({\hat p_1})\Delta^{\dagger}_{\beta \alpha_3}({\hat p_1}) \Delta_{\alpha_2 \gamma}({\hat p_2}) \Delta^{\dagger}_{\gamma \alpha_4}({\hat p_4}), \nonumber\\
{\rm Fig(e)}&=&\frac{1}{2}T\sum_{n_1}T\sum_{n_2}\int_{{\bf p}_1}\int_{{\bf p}_2}\int_{{\bf p}_3}G_{\varepsilon_1}({\bf p}_1)G_{-\varepsilon_1}(-{\bf p}_1)G_{\varepsilon_1}({\bf p}_3)G_{-\varepsilon_1}(-{\bf p}_3)G_{\varepsilon_2}({\bf p}_2)G_{-\varepsilon_2}(-{\bf p}_2)G_{\varepsilon_2}({\bf p}_4)G_{-\varepsilon_2}(-{\bf p}_4) \nonumber\\
&&\times |u({\bf p}_1-{\bf p}_3)|^2\Gamma^{(4)}_{\alpha_1,\alpha_2;\alpha_3,\alpha_4}({\bf p}_1,\varepsilon_1,{\bf p}_2,\varepsilon_2;{\bf p}_3,\varepsilon_1,{\bf p}_4,\varepsilon_2) \Delta_{\alpha_1 \beta}({\hat p_1})\Delta^{\dagger}_{\beta \alpha_3}({\hat p_3}) \Delta_{\alpha_2 \gamma}({\hat p_2}) \Delta^{\dagger}_{\gamma \alpha_4}({\hat p_4}), \nonumber\\
{\rm Fig(g)}&=&\frac{1}{2}T\sum_{n_1}T\sum_{n_2}\int_{{\bf p}_1}\int_{{\bf p}_2}\int_{{\bf p}_3}G^2_{\varepsilon_1}({\bf p}_1)G_{-\varepsilon_1}(-{\bf p}_1)G_{\varepsilon_1}({\bf p}_3)G^2_{\varepsilon_2}({\bf p}_2)G_{-\varepsilon_2}(-{\bf p}_2)G_{\varepsilon_2}({\bf p}_4) \nonumber\\
&&\times |u({\bf p}_1-{\bf p}_3)|^2\Gamma^{(4)}_{\alpha_1,\alpha_2;\alpha_3,\alpha_4}({\bf p}_1,\epsilon_1,{\bf p}_2,\epsilon_2;{\bf p}_3,\epsilon_1,{\bf p}_4,\epsilon_2) \Delta_{\alpha_1 \beta}({\hat p_1})\Delta^{\dagger}_{\beta \alpha_3}({\hat p_1}) \Delta_{\alpha_2 \gamma}({\hat p_2}) \Delta^{\dagger}_{\gamma \alpha_4}({\hat p_2}).  \nonumber
\end{eqnarray}

\begin{figure}[t]
\includegraphics[scale=0.47]{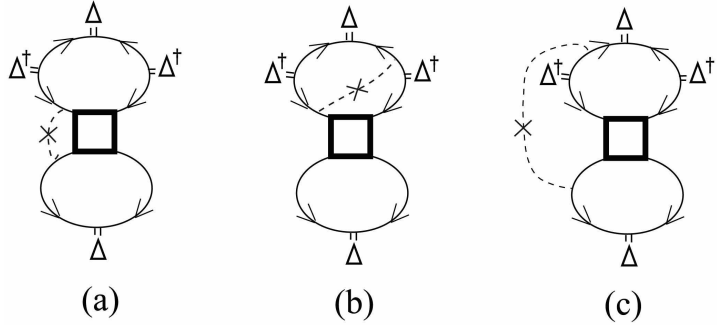}
\caption{Impurity-induced diagrams occurring from Fig.2A which can be neglected in the present analysis. \label{fig:neg}}
\end{figure} 

Hereafter, let us construct formulae of two terms, $\delta \beta_j$ and $\delta {\tilde \beta}_j$, giving impurity-dependent SC corrections to $\beta_j$ which are useful numerically once an appropriate FPV is substituted. Further, as far as the lowest order contributions in $T/E_F$ are concerned, one can neglect any frequency dependence of FPVs \cite{RS} in Fig.4, and the integral over $\xi_p= p^2/(2m) - \mu$ can be simplified using eq.(7). 
Next, the renormalized FPV, i.e.,  the scattering amplitude due to mutual interactions between the quasiparticles, will be decomposed \cite{AGD,SS} into a spin-symmetric part $T^{(s)}$ and a spin-antisymmetric part $T^{(a)}$
\begin{equation}\label{eq:amplitude}
2N(0)\Gamma^{(4)}_{\alpha\beta;\gamma\delta}=T^{\rm (s)}\delta_{\alpha\gamma}\delta_{\beta\delta}+T^{\rm (a)}(\vec{\sigma})_{\alpha\gamma}\cdot (\vec{\sigma})_{\beta\delta}. 
\end{equation}
Alternatively, it can also be expressed by the spin singlet amplitude $T_s$ and the triplet amplitude $T_t$ as 
\begin{equation}
2N(0)\Gamma^{(4)}_{\alpha\beta;\gamma\delta} = \frac{1}{2}T_s(\sigma_2)_{\alpha\beta}(\sigma_2)_{\delta\gamma} + \frac{1}{2}T_t(\sigma_2 \vec{\sigma})_{\alpha\beta} \cdot (\vec{\sigma} \sigma_2)_{\delta\gamma}, 
\end{equation}
where 
\begin{equation}
T_s=T^{\rm (s)}-3T^{\rm (a)}, \qquad T_t=T^{\rm (s)}+T^{\rm (a)}.
\end{equation}
Further, by assuming the bare impurity scattering amplitude ${\overline {|u({\bf p})|^2}}$ to be dominated by the $s$-wave component and separating the momentum integrals from the frequency summations, the diagrams in Fig.2 and Fig.4 are rewritten as    
\begin{eqnarray}\label{eq.all1}
{\rm Fig(C)}&=-&\frac{N(0)}{16 T^2}\tilde{S}_C\frac{T}{E_F}\int\frac{d\Omega_1}{4\pi}\int\frac{d\Omega_2}{4\pi}\int\frac{d\Omega_3}{4\pi}\delta(|\hat{p_1}+\hat{p_2}-\hat{p_3}|-1)\nonumber\\
&&\times \Big(\big[T^{\rm (s)}(\hat{p_1},\hat{p_2};\hat{p_3},\hat{p_4})\big]^2\big\{[\vec{\Delta}^*({\hat p_1})\cdot\vec{\Delta}(\hat{p_3})][\vec{\Delta}^*({\hat p_2})\cdot\vec{\Delta}(\hat{p_4})]\nonumber\\
&&\qquad\qquad\qquad\qquad -[\vec{\Delta}({\hat p_3})\cdot\vec{\Delta}(\hat{p_4})][\vec{\Delta}^*({\hat p_1})\cdot\vec{\Delta}^*(\hat{p_2})]+[\vec{\Delta}^*({\hat p_2})\cdot\vec{\Delta}(\hat{p_3})][\vec{\Delta}^*({\hat p_1})\cdot\vec{\Delta}(\hat{p_4})]\big\}\nonumber\\
&& +\big[T^{\rm (a)}(\hat{p_1},\hat{p_2};\hat{p_3},\hat{p_4})\big]^2\big\{-5[\vec{\Delta}^*({\hat p_1})\cdot\vec{\Delta}(\hat{p_3})][\vec{\Delta}^*({\hat p_2})\cdot\vec{\Delta}(\hat{p_4})]\nonumber\\
&&\qquad\qquad\qquad\qquad +5[\vec{\Delta}({\hat p_3})\cdot\vec{\Delta}(\hat{p_4})][\vec{\Delta}^*({\hat p_1})\cdot\vec{\Delta}^*(\hat{p_2})]+3[\vec{\Delta}^*({\hat p_2})\cdot\vec{\Delta}(\hat{p_3})][\vec{\Delta}^*({\hat p_1})\cdot\vec{\Delta}(\hat{p_4})]\big\}\Big), \nonumber\\
{\rm Fig(D)}&=&\frac{N(0)}{4 T^2}\tilde{S}_D\frac{T}{E_F}\int\frac{d\Omega_1}{4\pi}\int\frac{d\Omega_2}{4\pi}\int\frac{d\Omega_3}{4\pi}\delta(|\hat{p_1}+\hat{p_2}-\hat{p_3}|-1)\nonumber\\
&&\times \Big(\big[T^{\rm (s)}(\hat{p_1},\hat{p_2};\hat{p_3},\hat{p_4})T^{\rm (s)}(\hat{p_3},-\hat{p_2};\hat{p_1},-\hat{p_4})\nonumber\\
&&\qquad +T^{\rm (a)}(\hat{p_1},\hat{p_2};\hat{p_3},\hat{p_4})T^{\rm (a)}(\hat{p_3},-\hat{p_2};\hat{p_1},-\hat{p_4})\big]|\vec{\Delta}({\hat p_1})|^2[\vec{\Delta}^*({\hat p_2})\cdot\vec{\Delta}(\hat{p_4})]\nonumber\\
&&\qquad\quad\ +\big[T^{\rm (s)}(\hat{p_1},\hat{p_2};\hat{p_3},\hat{p_4})T^{\rm (a)}(\hat{p_3},-\hat{p_2};\hat{p_1},-\hat{p_4})+T^{\rm (a)}(\hat{p_1},\hat{p_2};\hat{p_3},\hat{p_4})T^{\rm (s)}(\hat{p_3},-\hat{p_2};\hat{p_1},-\hat{p_4})\big]\nonumber\\
&&\qquad\qquad\qquad\qquad \times\big\{[\vec{\Delta}({\hat p_1})\cdot\vec{\Delta}^*(\hat{p_2})][\vec{\Delta}^*({\hat p_1})\cdot\vec{\Delta}(\hat{p_4})]-[\vec{\Delta}({\hat p_1})\cdot\vec{\Delta}(\hat{p_4})][\vec{\Delta}^*({\hat p_1})\cdot\vec{\Delta}^*(\hat{p_2})]\big\}\Big), \nonumber\\
{\rm Fig(E)}&=-&\frac{N(0)}{16 T^2}\tilde{S}_E\frac{T}{E_F}\int\frac{d\Omega_1}{4\pi}\int\frac{d\Omega_2}{4\pi}\int\frac{d\Omega_3}{4\pi}\delta(|\hat{p_1}+\hat{p_2}-\hat{p_3}|-1)\nonumber\\
&&\times \Big(\big[T^{\rm (s)}(\hat{p_1},\hat{p_2};\hat{p_3},\hat{p_4})\big]^2\big[|\vec{\Delta}({\hat p_1})|^2|\vec{\Delta}(\hat{p_3})|^2-|\vec{\Delta}({\hat p_1})\cdot\vec{\Delta}(\hat{p_3})|^2+|\vec{\Delta}({\hat p_1})\cdot\vec{\Delta}^*(\hat{p_3})|^2\big]\nonumber\\
&& +\big[T^{\rm (a)}(\hat{p_1},\hat{p_2};\hat{p_3},\hat{p_4})\big]^2\big[3|\vec{\Delta}({\hat p_1})|^2|\vec{\Delta}(\hat{p_3})|^2+|\vec{\Delta}({\hat p_1})\cdot\vec{\Delta}(\hat{p_3})|^2-|\vec{\Delta}({\hat p_1})\cdot\vec{\Delta}^*(\hat{p_3})|^2\big]\Big), \nonumber
\end{eqnarray}
\begin{eqnarray}\label{eq.all2}
{\rm Fig(F)}&=-&\frac{N(0)}{32 T^2}\tilde{S}_F\frac{T}{E_F}\int\frac{d\Omega_1}{4\pi}\int\frac{d\Omega_2}{4\pi}\int\frac{d\Omega_3}{4\pi}\delta(|\hat{p_1}+\hat{p_2}-\hat{p_3}|-1)\nonumber\\
&&\times \Big(\big[T^{\rm (s)}(\hat{p_1},\hat{p_2};\hat{p_3},\hat{p_4})\big]^2\big[|\vec{\Delta}({\hat p_1})|^2|\vec{\Delta}(\hat{p_2})|^2-|\vec{\Delta}({\hat p_1})\cdot\vec{\Delta}(\hat{p_2})|^2+|\vec{\Delta}({\hat p_1})\cdot\vec{\Delta}^*(\hat{p_2})|^2\big]\nonumber\\
&& +\big[T^{\rm (a)}(\hat{p_1},\hat{p_2};\hat{p_3},\hat{p_4})\big]^2\big[3|\vec{\Delta}({\hat p_1})|^2|\vec{\Delta}(\hat{p_2})|^2+|\vec{\Delta}({\hat p_1})\cdot\vec{\Delta}(\hat{p_2})|^2-|\vec{\Delta}({\hat p_1})\cdot\vec{\Delta}^*(\hat{p_2})|^2\big]\Big), \nonumber\\
{\rm Fig(a)}&=&\frac{N(0)^2}{16\pi T^2}\Big(\sum_{n\geq 0} \frac{1}{(n+\frac{1}{2}+\frac{1}{4\pi\tau})^2}\Big)^2\frac{1}{\tau E_F}\int\frac{d\Omega_1}{4\pi}\int\frac{d\Omega_2}{4\pi}\int\frac{d\Omega_3}{4\pi}\delta(|\hat{p_1}+\hat{p_2}-\hat{p_3}|-1)\nonumber\\
&&\times \Big(T^{\rm (s)}(\hat{p_1},\hat{p_2};\hat{p_3},\hat{p_4})|\vec{\Delta}({\hat p_1})|^2|\vec{\Delta}({\hat p_4})|^2 \nonumber\\
&& +T^{\rm (a)}(\hat{p_1},\hat{p_2};\hat{p_3},\hat{p_4})\big\{|\vec{\Delta}({\hat p_1})\cdot \vec{\Delta}^*({\hat p_4})|^2-|\vec{\Delta}({\hat p_1})\cdot \vec{\Delta}({\hat p_4})|^2\big\}\Big), \nonumber\\
{\rm Fig(b)}&=-&\frac{N(0)^2}{8\pi T^2}\Big(\sum_{n\geq 0} \frac{1}{(n+\frac{1}{2}+\frac{1}{4\pi\tau})^2}\Big)^2\frac{1}{\tau E_F}\int\frac{d\Omega_1}{4\pi}\int\frac{d\Omega_2}{4\pi}\int\frac{d\Omega_3}{4\pi}\delta(|\hat{p_1}+\hat{p_2}-\hat{p_3}|-1)\nonumber\\
&&\times \Big(T^{\rm (s)}(\hat{p_1},\hat{p_2};\hat{p_3},\hat{p_4})\big[\vec{\Delta}({\hat p_1})\cdot\vec{\Delta}^*({\hat p_3})\big]|\vec{\Delta}({\hat p_4})|^2 \nonumber\\
&& +T^{\rm (a)}(\hat{p_1},\hat{p_2};\hat{p_3},\hat{p_4})\big\{\big[\vec{\Delta}({\hat p_1})\cdot \vec{\Delta}^*({\hat p_4})\big]\big[\vec{\Delta}({\hat p_4})\cdot \vec{\Delta}^*({\hat p_3})\big]-\big[\vec{\Delta}({\hat p_1})\cdot \vec{\Delta}({\hat p_4})\big]\big[\vec{\Delta}^*({\hat p_3})\cdot \vec{\Delta}^*({\hat p_4})\big]\big\}\Big), \nonumber\\
{\rm Fig(c)}&=&\frac{N(0)^2}{16\pi T^2}\Big(\sum_{n\geq 0} \frac{1}{(n+\frac{1}{2}+\frac{1}{4\pi\tau})^2}\Big)^2\frac{1}{\tau E_F}\int\frac{d\Omega_1}{4\pi}\int\frac{d\Omega_2}{4\pi}\int\frac{d\Omega_3}{4\pi}\delta(|\hat{p_1}+\hat{p_2}-\hat{p_3}|-1)\nonumber\\
&&\times \Big(T^{\rm (s)}(\hat{p_1},\hat{p_2};\hat{p_3},\hat{p_4})|\vec{\Delta}({\hat p_3})|^2|\vec{\Delta}({\hat p_4})|^2 \nonumber\\
&& +T^{\rm (a)}(\hat{p_1},\hat{p_2};\hat{p_3},\hat{p_4})\big\{|\vec{\Delta}({\hat p_3})\cdot \vec{\Delta}^*({\hat p_4})|^2-|\vec{\Delta}({\hat p_3})\cdot \vec{\Delta}({\hat p_4})|^2\big\}\Big), \nonumber\\
{\rm Fig(d)}&=-&\frac{N(0)^2}{8\pi T^2}\Big(\sum_{n\geq 0} \frac{1}{(n+\frac{1}{2}+\frac{1}{4\pi\tau})^2}\Big)^2\frac{1}{\tau E_F}\int\frac{d\Omega_1}{4\pi}\int\frac{d\Omega_2}{4\pi}\int\frac{d\Omega_3}{4\pi}\delta(|\hat{p_1}+\hat{p_2}-\hat{p_3}|-1)\nonumber\\
&&\times \Big(T^{\rm (s)}(\hat{p_1},\hat{p_2};\hat{p_3},\hat{p_4})\big[\vec{\Delta}({\hat p_2})\cdot\vec{\Delta}^*({\hat p_4})\big]|\vec{\Delta}({\hat p_1})|^2 \nonumber\\
&& +T^{\rm (a)}(\hat{p_1},\hat{p_2};\hat{p_3},\hat{p_4})\big\{\big[\vec{\Delta}({\hat p_1})\cdot \vec{\Delta}^*({\hat p_4})\big]\big[\vec{\Delta}({\hat p_2})\cdot \vec{\Delta}^*({\hat p_1})\big]-\big[\vec{\Delta}({\hat p_1})\cdot \vec{\Delta}({\hat p_2})\big]\big[\vec{\Delta}^*({\hat p_1})\cdot \vec{\Delta}^*({\hat p_4})\big]\big\}\Big), \nonumber\\
{\rm Fig(e)}&=&\frac{N(0)^2}{4\pi T^2}\Big(\sum_{n\geq 0} \frac{1}{(n+\frac{1}{2}+\frac{1}{4\pi\tau})^2}\Big)^2\frac{1}{\tau E_F}\int\frac{d\Omega_1}{4\pi}\int\frac{d\Omega_2}{4\pi}\int\frac{d\Omega_3}{4\pi}\delta(|\hat{p_1}+\hat{p_2}-\hat{p_3}|-1)\nonumber\\
&&\times \Big(T^{\rm (s)}(\hat{p_1},\hat{p_2};\hat{p_3},\hat{p_4})\big[\vec{\Delta}({\hat p_1})\cdot\vec{\Delta}^*({\hat p_3})\big]\big[\vec{\Delta}({\hat p_2})\cdot\vec{\Delta}^*({\hat p_4})\big] \nonumber\\
&& +T^{\rm (a)}(\hat{p_1},\hat{p_2};\hat{p_3},\hat{p_4})\big\{\big[\vec{\Delta}({\hat p_1})\cdot \vec{\Delta}^*({\hat p_4})\big]\big[\vec{\Delta}({\hat p_2})\cdot \vec{\Delta}^*({\hat p_3})\big]-\big[\vec{\Delta}({\hat p_1})\cdot \vec{\Delta}({\hat p_2})\big]\big[\vec{\Delta}^*({\hat p_3})\cdot \vec{\Delta}^*({\hat p_4})\big]\big\}\Big), \nonumber\\
{\rm Fig(g)}&=&\frac{N(0)^2}{16\pi T^2}\Big(\sum_{n\geq 0} \frac{1}{(n+\frac{1}{2}+\frac{1}{4\pi\tau})^2}\Big)^2\frac{1}{\tau E_F}\int\frac{d\Omega_1}{4\pi}\int\frac{d\Omega_2}{4\pi}\int\frac{d\Omega_3}{4\pi}\delta(|\hat{p_1}+\hat{p_2}-\hat{p_3}|-1)\nonumber\\
&&\times \Big(T^{\rm (s)}(\hat{p_1},\hat{p_2};\hat{p_3},\hat{p_4})|\vec{\Delta}({\hat p_1})|^2|\vec{\Delta}({\hat p_2})|^2 \nonumber\\
&& +T^{\rm (a)}(\hat{p_1},\hat{p_2};\hat{p_3},\hat{p_4})\big\{|\vec{\Delta}({\hat p_1})\cdot \vec{\Delta}^*({\hat p_2})|^2-|\vec{\Delta}({\hat p_1})\cdot \vec{\Delta}({\hat p_2})|^2\big\}\Big), 
\end{eqnarray}
where the factors such as ${\tilde S}_A$ denote results of summations over Matsubara frequencies; 
\begin{eqnarray}
\tilde{S}_{E}=\tilde{S}_{F}&=&(\pi k_B T)^4\sum_{\varepsilon_1}\sum_{\varepsilon_2}\sum_{\varepsilon_3}\frac{{\rm sgn}\varepsilon_1}{|\tilde{\varepsilon_1}|^2}\frac{{\rm sgn}\varepsilon_2}{|\tilde{\varepsilon_2}|^2}[{\rm sgn}\varepsilon_3][{\rm sgn}(\varepsilon_1+\varepsilon_2-\varepsilon_3)], \nonumber\\
\tilde{S}_D &=&(\pi k_B T)^4\sum_{\varepsilon_1}\sum_{\varepsilon_2}\sum_{\varepsilon_3}\frac{1}{|\tilde{\varepsilon_1}|}\frac{1}{|\tilde{\varepsilon_2}|}\frac{{\rm sgn}\varepsilon_3}{|\tilde{\varepsilon_3}|^2}[{\rm sgn}(\varepsilon_1+\varepsilon_2-\varepsilon_3)], \nonumber\\
\tilde{S}_C&=&(\pi k_B T)^4\sum_{\varepsilon_1}\sum_{\varepsilon_2}\sum_{\varepsilon_3}\frac{1}{|\tilde{\varepsilon_1}|}\frac{1}{|\tilde{\varepsilon_2}|}\frac{1}{|\tilde{\varepsilon_3}|}\frac{1}{|\tilde{\varepsilon}_1+\tilde{\varepsilon}_2-\tilde{\varepsilon}_3|}. \nonumber
\end{eqnarray}

The remainder of our calculation is to perform angle averages over the Fermi surface. To conveniently parameterize $T^{\rm (s)}(\hat{p_1},\hat{p_2};\hat{p_3},\hat{p_4})$ and $T^{(a)}(\hat{p_1},\hat{p_2};\hat{p_3},\hat{p_4})$, which are functions of only two independent variables, we introduce Abrikosov-Khalatnikov angles $\theta$ and $\phi$, which are related to the angles between incident two particles or between an incident and an outgoing particles in the manner 
\begin{eqnarray}
\hat{p_1}\cdot\hat{p_2}=\hat{p_3}\cdot\hat{p_4}=\cos\theta\equiv x_1, 
\nonumber\\
\hat{p_1}\cdot\hat{p_3}=\cos^2\frac{\theta}{2}+\sin^2\frac{\theta}{2}\cos\phi\equiv x_2, \nonumber\\
\hat{p_1}\cdot\hat{p_4}=\cos^2\frac{\theta}{2}-\sin^2\frac{\theta}{2}\cos\phi\equiv x_3 .\nonumber
\end{eqnarray}
To treat the angle-integrals in eq.(11) including a $\delta$ function, we choose $\hat{p_1}+\hat{p_2}$ as the polar axis for $\hat{p_3}$, and the azimuthal angle $\phi$ of $\hat{p_3}$ will be defined by measuring it from the plane containing $\hat{p_1}$ and $\hat{p_2}$. Then, $d\Omega_3=d\cos\theta_3d\phi$, and we can perform the integral with respect to $\theta_3$. Further, the unit vector $\hat{p}$ will be chosen along $\hat{p_1}+\hat{p_2}$, and an additional azimuthal angle $\psi$ is chosen as an angle measured, for the new polar axis $\hat{p}$, from the plane containing $\hat{p_1}$ and $\hat{p_2}$. Then, we have \cite{RS} 
\[
\int\frac{d\Omega_1}{4\pi}\int\frac{d\Omega_2}{4\pi}\int\frac{d\Omega_3}{4\pi}\delta(|\hat{p_1}+\hat{p_2}-\hat{p_3}|-1)=\frac{1}{2}\int^1_0d\cos(\theta/2)\int^{2\pi}_0\frac{d\phi}{2\pi}\int\frac{d\Omega_p}{4\pi}\int^{2\pi}_0\frac{d\psi}{2\pi} .
\]
For instance, the product of scattering amplitudes $T(\hat{p_1},\hat{p_2};\hat{p_3},\hat{p_4})T(\hat{p_3},-\hat{p_2};\hat{p_1},-\hat{p_4})$ appearing in the expression for Fig.(D) in eq.(\ref{eq.all2}) can be expressed as $T(\theta,\phi)T(\theta ',\phi ')$, where $\theta '$ and $\phi '$ are related to $\theta$ and $\phi$ through the formula
\begin{eqnarray}
\cos \theta ' &=&\cos\phi-\cos^2(\theta/2)(\cos\phi+1), \nonumber\\
\cos\phi '&=& \frac{3\cos^2(\theta/2)-1-[\cos^2(\theta/2)-1]\cos\phi}{\cos^2(\theta/2)+1+[\cos^2(\theta/2)-1]\cos\phi}. \nonumber
\end{eqnarray}
For later convenience, the variables $x'_1$, $x'_2$, and $x'_3$ will also be defined as
\begin{eqnarray}
x'_1&\equiv&\cos\theta' =-x_3, \nonumber\\
x'_2&\equiv&\cos^2\frac{\theta'}{2}+\sin^2\frac{\theta'}{2}\cos\phi'=x_2, \nonumber\\
x'_3&\equiv&\cos^2\frac{\theta'}{2}-\sin^2\frac{\theta'}{2}\cos\phi'=-x_1. \nonumber
\end{eqnarray}
What we need to perform finally is to calculate the integral 
\begin{equation}
\int\frac{d\Omega_p}{4\pi}\int\frac{d\psi}{2\pi}[\vec{\Delta}^*(\hat{p_i})\cdot\vec{\Delta}(\hat{p_j})][\vec{\Delta}^*(\hat{p_k})\cdot\vec{\Delta}(\hat{p_l})].
\end{equation}

Since the angles $\theta$ and $\phi$ fix the relative position of the four vectors $\hat{p_i}(i=1-4)$, and the absolute positions of those vectors are determined by $\hat{p}$ and $\psi$, the polar coodinates $(\bar{\theta_i},\bar{\phi_i})$ of $\hat{p_i}$ are $(\bar{\theta_1},\bar{\phi_1})=(\theta/2,0)$, $(\bar{\theta_2},\bar{\phi_2})=(\theta/2,\pi)$, $(\bar{\theta_3},\bar{\phi_3})=(\theta,\phi)$, and $(\bar{\theta_4},\bar{\phi_4})=(\theta/2,\phi+\pi)$ in the body-fixed frame. Each component of $\vec{\Delta}(\hat{p_i})$ can be expanded in terms of $l=1$ spherical harmonics as 
\[
\Delta_{\mu}(\hat{p_i})=\sum^{l}_{m=-l}B_{\mu ,m}Y_{l m}(\hat{p_i}) 
\]
with $l=1$, where, by definition of $A_{\mu,j}$, the relations between $A_{\mu,j}$ and $B_{\mu,j}(l=1)$
\begin{eqnarray}\label{eq.relation}
B_{\mu,1}&=&-\frac{1}{2}\sqrt{\frac{8\pi}{3}}(A_{\mu,x}-iA_{\mu,y}),\nonumber\\
B_{\mu,-1}&=&\frac{1}{2}\sqrt{\frac{8\pi}{3}}(A_{\mu,x}+iA_{\mu,y}),\nonumber\\
B_{\mu,0}&=&\frac{1}{\sqrt{2}}\sqrt{\frac{8\pi}{3}}A_{\mu,z}
\end{eqnarray}
are satisfied. Then, each integral appearing in the expressions of eq.(11) 
reduces to 
\begin{equation}
\int\frac{d\Omega_p}{4\pi}\int\frac{d\psi}{2\pi} \, Y_{lm_1}(\hat{p_i})Y_{lm_2}(\hat{p_j})Y_{lm_3}(\hat{p_k})Y_{lm_4}(\hat{p_l}).
\end{equation} 
To perform these angle integrals, it is convenient to use the formula 
\[
Y_{lm}(\hat{p_i})=\sum^{l}_{m'=l}D^{(l)}_{mm'}(R)*Y_{lm'}(\bar{\theta_i},\bar{\phi_i}),
\]
where $R$ is the rotation which maps the coordinate system $(p_x,p_y,p_z)$ into the body coordinate system parameterized by the angles $\bar{\theta_i}$ and $\bar{\phi_i}$, and $D^{(l)}_{mm'}(R)$ is the corresponding rotation matrix. Using the standard properties of rotation matrices together with this transformation, we can 
obtain    
\begin{eqnarray}\label{eq.final}
\int\frac{d\Omega_p}{4\pi}\int\frac{d\psi}{2\pi}Y_{lm_1}(\hat{p_i})Y_{lm_2}(\hat{p_j})Y_{lm_3}(\hat{p_k})Y_{lm_4}(\hat{p_l}) &=& \sum^{2l}_{L=0}\frac{(-1)^{m_1+m_2}}{2L+1} \nonumber \\
&\times& \langle lm_1 l m_2|Lm_1+m_2\rangle \langle lm_3 l m_4|L-m_1-m_2\rangle \Phi^{(l)}_L (\theta,\phi),
\end{eqnarray}
where 
\begin{eqnarray}
\Phi^{(l)}_L (\theta,\phi)=\sum_{m'_1\cdots m'_4}(-1)^{m'_1+m'_2}\langle Lm'_1+m'_2|lm'_1 l m'_2\rangle \langle L-m'_1-m'_2|lm'_3 l m'_4\rangle Y_{lm'_1}(\bar{\theta_i},\bar{\phi_i})Y_{lm'_2}(\bar{\theta_j},\bar{\phi_j})Y_{lm'_3}(\bar{\theta_k},\bar{\phi_k})Y_{lm'_4}(\bar{\theta_l},\bar{\phi_l}).\nonumber
\end{eqnarray}  
Note that the Clebsch-Gordan coefficients in eq.(\ref{eq.final}) require $m_1+m_2+m_3+m_4=0$ and $m'_1+m'_2+m'_3+m'_4=0$. Using these expressions, one can obtain the contributions to the quartic term in the GL free energy functional. The same procedure has been used in clean limit in Ref.\cite{RS}.

Results of the SC correction to $\beta_j$ in the relaxation time approximation, given by (C)-(F) in Fig.2, are given by 
\begin{equation}
\delta\beta_j=\delta\beta^{\rm C}_j+\delta\beta^{\rm E+F}_j+\delta\beta^{\rm D}_j,
\end{equation}
where
\[
\delta\beta^{\rm K}_j=- \frac{4\pi^2}{7\zeta(3)} \frac{T}{E_F} \beta_0(T)(\tilde{S_{\rm K}}/16)\langle W^{\rm K}_j(\theta,\phi)[T^{(s)}(\theta,\phi)]^2+V^{\rm K}_j(\theta,\phi)[T^{(a)}(\theta,\phi)]^2\rangle
\]
for ${\rm K}={\rm C,E+F}$, where ${\tilde S}_{{\rm E}+{\rm F}} = {\tilde S}_{\rm E}={\tilde S}_{\rm F}$, and
\begin{eqnarray}\label{eq:sc}
\delta\beta^{\rm D}_j&=& - \frac{4\pi^2}{7\zeta(3)} \frac{T}{E_F} \beta_0(T)(\tilde{S_{\rm D}}/4)\langle W^{\rm D}_j(\theta,\phi)[T^{(s)}(\theta,\phi)T^{(s)}(\theta ',\phi ')+T^{(a)}(\theta,\phi)T^{(a)}(\theta ',\phi ')]\nonumber\\
&&\qquad\qquad\qquad\qquad +V^{\rm D}_j(\theta,\phi)[T^{(s)}(\theta,\phi)T^{(a)}(\theta ',\phi ')+T^{(a)}(\theta,\phi)T^{(s)}(\theta ',\phi ')]\rangle .
\end{eqnarray}
The weighting functions $W^{\rm K}_j$ and $V^{\rm K}_j$ are given in Table.I, and $\langle Z \rangle$ denotes the angular average $\int^1_0 d(\cos(\theta/2))\int^{2\pi}_0 \, Z \, d\phi/(2\pi)$.

\begin{table}[h]
\caption{Weighting functions for $\delta\beta_j^{\rm K}$.}
\begin{tabular}{>{$}c<{$}>{$}c<{$}>{$}c<{$}>{$}c<{$}}
\hline
\hline
\quad & {\rm K}={\rm C} & {\rm K}={\rm D} & {\rm K}={\rm E}+{\rm F}  \\
\hline
W^{({\rm K})}_1 & -4x^2_1+x^2_2+x^2_3 & 0 & -\frac{3}{2}x^2_1-3x^2_2+\frac{3}{2} \\
V^{({\rm K})}_1 &  20x^2_1-5x^2_2-5x^2_3& 3x_1x_3-x_2 & \frac{15}{2}x^2_1+3x^2_2-\frac{7}{2} \\
W^{({\rm K})}_2 & -2x^2_1+3x^2_2+3x^2_3 & 2x_1x_3-4x_2 & \frac{1}{2}x^2_1+x^2_2+\frac{9}{2} \\
V^{({\rm K})}_2 & 2x^2_1-23x^2_2+17x^2_3 & -3x_1x_3+x_2 & -\frac{21}{2}x^2_1-9x^2_2+\frac{43}{2} \\
W^{({\rm K})}_3 & 8x^2_1-2x^2_2-2x^2_3 & -3x_1x_3+x_2 & 3x^2_1+6x^2_2-3 \\
V^{({\rm K})}_3 & -8x^2_1+2x^2_2+2x^2_3 & -3x_1x_3+x_2 & -3x^2_1+6x^2_2-1 \\
W^{({\rm K})}_4 & -2x^2_1+3x^2_2+3x^2_3 & -3x_1x_3+x_2 & \frac{1}{2}x^2_1+x^2_2+\frac{9}{2} \\
V^{({\rm K})}_4 & 2x^2_1+17x^2_2-23x^2_3 & 2x_1x_3-4x_2 & \frac{19}{2}x^2_1+11x^2_2-\frac{37}{2} \\
W^{({\rm K})}_5 & 2x^2_1-3x^2_2-3x^2_3 & 0 & -\frac{1}{2}x^2_1-x^2_2-\frac{9}{2} \\
V^{({\rm K})}_5 & -10x^2_1+15x^2_2+15x^2_3 & x_1x_3+3x_2 & \frac{5}{2}x^2_1+x^2_2+\frac{21}{2} \\
\hline
\hline
\end{tabular}
\end{table}

On the other hand, the corresponding results of the SC corrections $\delta {\tilde \beta}_i$ including the impurity-induced vertex correction, which arise from (a)-(i) of Fig.4, are 
\begin{eqnarray}
\delta{\tilde \beta}_i^{\rm (a+i)}&=&\frac{2\pi^2}{7\zeta(3)}\beta_0(T)\frac{1}{4\pi\tau T}\frac{T}{E_F}\Big(\psi^{(1)}\big(\frac{1}{2}+\frac{1}{4\pi\tau T}\big)\Big)^2\langle W^{\rm (a)}_i T^{\rm (s)}(\theta,\phi)+V^{\rm (a)}_i T^{\rm (a)}(\theta,\phi)\rangle, \nonumber\\
\delta{\tilde \beta}_i^{\rm (b+f)}&=&-\frac{4\pi^2}{7\zeta(3)}\beta_0(T)\frac{1}{4\pi\tau T}\frac{T}{E_F}\Big(\psi^{(1)}\big(\frac{1}{2}+\frac{1}{4\pi\tau T}\big)\Big)^2\langle W^{\rm (b)}_i T^{\rm (s)}(\theta,\phi)+V^{\rm (b)}_i T^{\rm (a)}(\theta,\phi)\rangle, \nonumber\\
\delta{\tilde \beta}_i^{\rm (c)}&=&\frac{\pi^2}{7\zeta(3)}\beta_0(T)\frac{1}{4\pi\tau T}\frac{T}{E_F}\Big(\psi^{(1)}\big(\frac{1}{2}+\frac{1}{4\pi\tau T}\big)\Big)^2\langle W^{\rm (c)}_i T^{\rm (s)}(\theta,\phi)+V^{\rm (c)}_i T^{\rm (a)}(\theta,\phi)\rangle, \nonumber\\
\delta{\tilde \beta}_i^{\rm (d+h)}&=&-\frac{4\pi^2}{7\zeta(3)}\beta_0(T)\frac{1}{4\pi\tau T}\frac{T}{E_F}\Big(\psi^{(1)}\big(\frac{1}{2}+\frac{1}{4\pi\tau T}\big)\Big)^2\langle W^{\rm (d)}_i T^{\rm (s)}(\theta,\phi)+V^{\rm (d)}_i T^{\rm (a)}(\theta,\phi)\rangle, \nonumber\\
\delta{\tilde \beta}_i^{\rm (e)}&=&\frac{4\pi^2}{7\zeta(3)}\beta_0(T)\frac{1}{4\pi\tau T}\frac{T}{E_F}\Big(\psi^{(1)}\big(\frac{1}{2}+\frac{1}{4\pi\tau T}\big)\Big)^2\langle W^{\rm (e)}_i T^{\rm (s)}(\theta,\phi)+V^{\rm (e)}_i T^{\rm (a)}(\theta,\phi)\rangle, \nonumber\\
\delta{\tilde \beta}_i^{\rm (g)}&=&\frac{\pi^2}{7\zeta(3)}\beta_0(T)\frac{1}{4\pi\tau T}\frac{T}{E_F}\Big(\psi^{(1)}\big(\frac{1}{2}+\frac{1}{4\pi\tau T}\big)\Big)^2\langle W^{\rm (g)}_i T^{\rm (s)}(\theta,\phi)+V^{\rm (g)}_i T^{\rm (a)}(\theta,\phi)\rangle, \nonumber
\end{eqnarray}
where the weighting functions $W_i^{({\rm k})}$ and $V_i^{({\rm k})}$ are shown in Table.II. 
\begin{table}[h]
\caption{Weighting functions for $\delta {\tilde \beta}_j^{(k)}$.}
\begin{tabular}{>{$}c<{$}>{$}c<{$}>{$}c<{$}>{$}c<{$}>{$}c<{$}}
\hline
\hline
\quad & {\rm k}={\rm a} & {\rm k}={\rm b}={\rm d} & {\rm k}={\rm c}={\rm g} & {\rm k}={\rm e} \\
\hline
W^{({\rm k})}_1 & 0 & 0 & 0 & 0 \\
V^{({\rm k})}_1 & -3x_3^2+1 & -3x_1x_3+x_2 & -3x_1^2+1 & -4x_1^2+x_2^2+x_3^2 \\
W^{({\rm k})}_2 & -2x_3^2+4 & 4x_2-2x_1x_3 & -2x_1^2+4 & -x_1^2+4x_2^2-4x_3^2 
\\
V^{({\rm k})}_2 & 3x_3^2-1 & 3x_1x_3-x_2 & 3x_1^2-1 & -x_1^2-x_2^2+4x_3^2 \\
W^{({\rm k})}_3 & 3x_3^2-1 & 3x_1x_3-x_2 & 3x_1^2-1 & 4x_1^2-x_2^2-x_3^2 \\
V^{({\rm k})}_3 & 3x_3^2-1 & 3x_1x_3-x_2 & 3x_1^2-1 & 4x_1^2-x_2^2-x_3^2 \\
W^{({\rm k})}_4 & 3x_3^2-1 & 3x_1x_3-x_2 & 3x_1^2-1 & -x_1^2-x_2^2+4x_3^2 \\
V^{({\rm k})}_4 & -2x_3^2+4 & -2x_1x_3+4x_2 & -2x_1^2+4 & -x_1^2+4x_2^2-x_3^2 
\\
W^{({\rm k})}_5 & 0 & 0 & 0 & 0 \\
V^{({\rm k})}_5 & -x_3^2-3 & -x_1x_3-3x_2 & -x_1^2-3 & 2x_1^2-3x_2^2-3x_3^2 \\
\hline
\hline
\end{tabular}
\end{table}

 Summarizing the above results, we reach the formula of $\delta {\tilde \beta}_j$ 
\begin{equation}\label{eq:strongimp}    
\delta\tilde{\beta}_i = - \frac{4\pi^2}{7\zeta(3)}\beta_0(T)\frac{1}{4\pi\tau T}\frac{T}{E_F}\Big(\psi^{(1)}\big(\frac{1}{2}+\frac{1}{4\pi\tau T}\big)\Big)^2\langle W^{\rm (total)}_i T^{\rm (s)}(\theta,\phi)+V^{\rm (total)}_i T^{\rm (a)}(\theta,\phi)\rangle, 
\end{equation}
which represents the new impurity-induced SC correction and should be added to $\delta \beta_j$. Here, the values of $W^{\rm total}_i$ and $V^{\rm total}_i$ are shown in Table.III.
\begin{table}[h]
\begin{center}
\caption{Weighting functions appeared in eq.(18).}
\begin{tabular}{>{$}c<{$}>{$}c<{$}}
\hline
\hline
W^{\rm (total)}_1 & 0 \\
V^{\rm (total)}_1 & \frac{9}{2}x_1^2-\frac{1}{2}x_3^2-4x_1x_3 \\
W^{\rm(total)}_2 & -2x_1^2-2x_3^2+4x_1x_3 \\
V^{\rm (total)}_2 & \frac{1}{2}x_1^2-\frac{9}{2}x_3^2+4x_1x_3 \\
W^{\rm (total)}_3 & -\frac{9}{2}x_1^2+\frac{1}{2}x_3^2+4x_1x_3 \\
V^{\rm (total)}_3 & -\frac{9}{2}x_1^2+\frac{1}{2}x_3^2+4x_1x_3 \\
W^{\rm (total)}_4 & \frac{1}{2}x_1^2-\frac{9}{2}x_3^2+4x_1x_3 \\
V^{\rm (total)}_4 & -2x_1^2-2x_3^2+4x_1x_3 \\
W^{\rm (total)}_5 & 0 \\
V^{\rm (total)}_5 & \frac{3}{2}x_1^2+\frac{13}{2}x_3^2-8x_1x_3 \\
\hline
\hline
\end{tabular}
\end{center}
\end{table}
\\

Our remaining task is to determine the FPV concretely. To do this we use two approaches; one is a perturbative approach starting from a bare repulsive interaction \cite{rep} and the other is a phenomenological approach in which the FPV is determined in terms of Landau parameters estimated from experimental data of transport coefficients and others \cite{SS}.

\subsection{Perturbative approach}

In this approach \cite{rep}, the attractive interaction between the quasiparticles is assumed to be induced by a short-range repulsive interaction between two bare particles, and hence, we start from the following two particle interaction term 
\begin{equation}
H_{\rm int}=\frac{g}{2}\sum_{\alpha}\int_{\bf r}\int_{\bf r'}\Psi^{\dagger}_{\alpha}({\bf r})\Psi^{\dagger}_{-\alpha}({\bf r'}) \, \delta({\bf r}-{\bf r'}) \, \Psi_{-\alpha}({\bf r})\Psi_{\alpha}({\bf r'}), 
\end{equation} 
where $g$ is a positive coupling constant. We carry out the perturbative expansion in $g$ and calculate the FPV function $\Gamma^{(4)}_{\alpha\beta;\gamma\delta}$ up to the second order in $g$. 
\begin{eqnarray}
\Gamma^{(4)}_{\alpha\beta;\gamma\delta}({\bf p}_1,{\bf p}_2;{\bf p}_3,{\bf p}_1+{\bf p}_2-{\bf p}_3)&=&\big[\frac{1}{2}g+2g^2\Pi({\bf p}_2-{\bf p}_3)+g^2C({\bf p}_1+{\bf p}_2)-g^2\Pi({\bf p}_1-{\bf p}_3)\big]\delta_{\alpha\gamma}\delta_{\beta\delta}\nonumber\\
&&\qquad\qquad -\big[\frac{1}{2}g+g^2C({\bf p}_1+{\bf p}_2)+g^2\Pi({\bf p}_1-{\bf p}_3)\big](\vec{\sigma})_{\alpha\gamma}\cdot(\vec{\sigma})_{\beta\delta}, 
\end{eqnarray}
where \cite{rep} 
\[
\Pi({\bf q})=\frac{1}{2}N(0)\Big[1+\frac{4p^2_F-{\bf q}^2}{4p_F|{\bf q}|}\ln\frac{2p_F+|{\bf q}|}{2p_F-|{\bf q}|}\Big]
\]
is a polarization function,
\[
C({\bf q})=N(0)\Big[1+\frac{\sqrt{1-{\bf q}^2/4p^2_F}}{2}\ln\frac{\sqrt{1-{\bf q}^2/4p^2_F}-1}{\sqrt{1-{\bf q}^2/4p^2_F}+1}\Big]
\]
is a Cooper loop regularized at a large momentum. Comparing with Eq.(\ref{eq:amplitude}) we can obtain $T^{\rm (s)}$ and $T^{\rm (a)}$ with the use of the Abrikosov-Khalatnikov angles $\theta$ and $\phi$ in the manner 
\begin{eqnarray}
T^{\rm (s)}(\theta,\phi)&=&\lambda+\lambda^2\{2\pi(x_3)+c(x_1)-\pi(x_2)\}, \nonumber\\
T^{\rm (a)}(\theta,\phi)&=&-\lambda-\lambda^2\{c(x_1)+\pi(x_2)\}, 
\end{eqnarray}
where $\lambda=gN(0)$, and the functions $\pi(x)$ and $c(x)$, corresponding to $\Pi({\bf q})$ and $C({\bf q})$, respectively, are given by
\begin{eqnarray}
\pi(x) &=& 1+\frac{\sqrt{2}(1+x)}{4\sqrt{1-x}}\ln \frac{\sqrt{2}+\sqrt{1-x}}{\sqrt{2}-\sqrt{1-x}}, \nonumber \\
c(x) &=& 2\Big\{1+\frac{\sqrt{1-x}}{2\sqrt{2}}\ln \frac{\sqrt{2}-\sqrt{1-x}}{\sqrt{2}+\sqrt{1-x}}\Big\}.
\end{eqnarray}

As one can see from the formula (\ref{eq:sc}), the combination of $T^{(i)}T^{(j)}(i,j=s,a)$ always appears in calculating the SC corrections to $\beta_j$. Hence, we can include third-order corrections in $\lambda$ to the quartic term of the GL free energy functional. The products of two vertex parts can be expressed as 
\begin{eqnarray}
\big[T^{\rm (s)}(\theta,\phi)\big]^2&=&\lambda^2+2\lambda^3 \{2\pi(x_3)+c(x_1)-\pi(x_2)\}, \nonumber\\ 
\big[T^{\rm (a)}(\theta,\phi)\big]^2&=&\lambda^2+2\lambda^3 \{c(x_1)+\pi(x_2)\}, \nonumber\\
T^{\rm (s)}(\theta,\phi)T^{\rm (a)}(\theta',\phi')&=&-\lambda^2-\lambda^3\{2\pi(x_3)+c(x_1)-\pi(x_2)+c(x'_1)+\pi(x'_2)\}, \nonumber\\
T^{\rm (a)}(\theta,\phi)T^{\rm (s)}(\theta',\phi')&=&-\lambda^2-\lambda^3\{c(x_1)+\pi(x_2)+2\pi(x'_3)+c(x'_1)-\pi(x'_2)\}. 
\end{eqnarray}

Then, $\delta\beta_j$, i.e., the SC contribution to $\beta_j$ in the relaxation time approximation, is given by 
\begin{eqnarray}\label{eq.rtabetaj}
\delta\beta_j &=& - \beta_0(T)\frac{T}{E_F}\lambda^2\Big[\big(m_j^{\rm C}\tilde{S}_C+m_j^{\rm E+F}\tilde{S}_{E+F}+m_j^{\rm D}\tilde{S}_D\big) \nonumber \\
&+& \lambda \big(n_j^{\rm C}\tilde{S}_C+n_j^{\rm E+F}\tilde{S}_{E+F}+n_j^{\rm D}\tilde{S}_D\big) \Big],
\end{eqnarray}
where the coefficients in eq.(\ref{eq.rtabetaj}) are given in Table IV. 
\begin{table}[h]
\begin{center}
\caption{Numerical values of $m_j^{\rm K}$ and $n_j^{\rm K}$ in eq.(\ref{eq.rtabetaj}).}
\begin{tabular}{c  c c  c c c c }
\hline\hline
j& $m_j^{\rm C}$    & $m_j^{\rm E+F}$ & $m_j^{\rm D}$ & $n_j^{\rm C}$ & $n_j^{\rm E+F}$ &$n_j^{\rm D}$ \\
\hline
\\
1  & 1.0943  & 0.2344 & 0.3128 & -4.0586 & -1.4992 & -0.1950 \\
2  & 0.0002 & 5.1601 & -3.1273 & -0.7179 & 12.3408 & -12.0611 \\
3  & 0  & 0.4689 & 0 & 0 & 1.1323 & 0 \\
4  & -0.0002 & 1.0948 & 3.1273 & 0.7180 & -4.8988 & 12.0611 \\
5  & 2.1890  & 2.0326 & -2.5016 & 7.8083 & 3.0325 & -12.4511 \\
\\
\hline\hline
\end{tabular}
\end{center}
\end{table}

As the limit of vanishing $\tau^{-1}$ of eq.(22), the corresponding expression in {\it clean limit} becomes
\begin{equation}
\delta\beta_j =-\beta_0(T)\frac{T}{2E_F}\lambda^2\big(M_j+\lambda N_j \big), 
\end{equation}
where $M_1=76.1$, $N_1=-271.3$, $M_2=7.2$, $N_2=-119.6$, $M_3=6.4$, $N_3=15.5$, $M_4=48.4$, $N_4=221.5$, $M_5=110.3$, and $N_5=264.1$. It is found that the use of these numerical values of coefficients leads to reproducing the numerical results in Ref.\cite{rep}. For the ABM pairing state, the coefficients are given by $M_{\rm ABM}=M_{245}=165.9$ and $N_{\rm ABM}=366.0$, while those for the BW pairing state are $M_{\rm BW}=M_{12}+M_{345}/3=138.3$ and $N_{\rm BW}=-223.9$. The third order terms are important for the stability of the ABM pairing state because the sign of $N_{\rm BW}$ is negative contrary to that of $N_{\rm ABM}$. 

On the other hand, the result on eq.(\ref{eq:strongimp}) in this approach is given by
\begin{equation}
\delta\tilde{\beta}_j=-\frac{4\pi^2}{7\zeta(3)}\beta_0(T)\frac{\lambda}{4\pi\tau T}\frac{T}{E_F}\Big(\psi^{(1)}\big(\frac{1}{2}+\frac{1}{4\pi\tau T}\big)\Big)^2\big(P_j+\lambda Q_j \big), 
\end{equation}
where $P_1=-0.81$, $Q_1=1.4$, $P_2=0$, $Q_2=0.002$, $P_3=0$, $Q_3=-0.66$, $P_4=0$, $Q_4=-1.8$, $P_5=-1.6$ and $Q_5=0.45$. That is, $P_{\rm ABM}=-1.6$ and $Q_{\rm ABM}=-1.32$ while $P_{\rm BW}=-1.3$ and $Q_{\rm BW}=0.81$. The third order term in both $\lambda$ and $1/\tau$ has, in the BW pairing state, the opposite sign to that of the ABM pairing one. That is, the impurity effects in the BW state enhance the SC correction, while they rather reduce that in the ABM pairing 
state. 

\begin{figure}[h]
\includegraphics[scale=0.24]{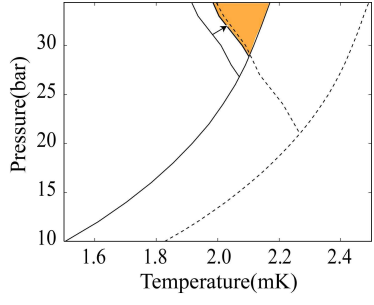}
\caption{Mean field phase diagram of the model with no anisotropy obtained in the perturbative approach and in terms of $1/2\pi\tau =0.13$ (mK) and $\lambda=0.821$. The dashed lines are the transition curves of bulk $^3$He. By including the impurity-induced contribution, eq.(\ref{eq:strongimp}), the stability of the ABM pairing state is weakened and, as indicated by the arrow, its region shrinks until reaching the shaded region. \label{fig:phaserep}}  
\end{figure}

A typical result of T-P phase diagram in this perturbative approach is shown in Fig.\ref{fig:phaserep} where $\lambda$ was assumed, for simplicity, to be independent of pressure. In obtaining a phase diagram based on our calculations, we use hereafter the experimental $P$-dependence \cite{VW} of the bulk transition temperature $T_{c0}(P)$ and the $P$-dependence of the Fermi energy given in Ref.\cite{Wheatley}. The dashed lines are the transition curves in clean limit (i.e., of the bulk liquid). The resulting bulk $T_{AB}(P)$-curve is slightly wavy because the neglect of $P$-dependence of $\lambda$ and the perturbative treatment in $\lambda$ may not be fully justified. The lower solid curve with negative $dP/dT$ values is the $T_{AB}$-line obtained in the relaxation time approximation, i.e., by neglecting $\delta\tilde{\beta}_j$. On the other hand, the upper (inner) solid curve close to the corresponding (dashed) curve of the bulk liquid is the $T_{AB}$ curve resulting from the full expressions including $\delta\tilde{\beta}_j$. Here, we have chosen the value $1/(2\pi\tau)=0.13$ (mK). From the figure, one can see that the impurity-induced SC correction, $\delta\tilde{\beta}_j$, weakenes the stability of the ABM pairing state.\\        

\subsection{Phenomenological approach}

In this approach \cite{SS}, the quasiparticle FPV function $\Gamma^{(4)}(\hat{p_1},\hat{p_2};\hat{p_3},\hat{p_4})$ depends on two momentum transfers $q_1=|\hat{p_1}-\hat{p_3}|=2k_F\sqrt{(1-x_2)/2}$ and $q_2=|\hat{p_1}-\hat{p_4}|=2k_F\sqrt{(1-x_3)/2}$. The singlet and triplet scattering amplitudes are approximated by $
T_s(x_2,x_3)=W_s(x_2) + W_s(x_3)$ and $T_t(x_2,x_3)=W_t(x_2)-W_t(x_3)$, 
which automatically satisfy the exchange antisymmetry conditions, $T_s(x_2,x_3)=T_s(x_3,x_2)$ and $T_t(x_2,x_3)=-T_t(x_3,x_2)$. We expand $W_s$ and $W_t$ in terms of the partial waves 
\begin{eqnarray}
W_s(x) &=& \sum^{\infty}_{l=0}W^s_lP_l(x), \nonumber \\
W_t(x) &=& \sum^{\infty}_{l=0}W^t_lP_l(x), 
\end{eqnarray}
where $P_l(x)$ is a Legendre polynomial, and determine the coefficients $W^{s,t}_l$ from experimental data. Note that $W^t_0$ can be ignored in considering the physical quantities since $l=0$ component disappears from $T_t(x_2,x_3)$. Following Ref.\cite{SS} we use the quasiparticle lifetime, the thermal conductivity, the viscosity, and the spin diffusion to determine three Landau parameters $F_0^s$, $F_0^a$, and $F_1^s$ through experimental estimations for them and to obtain the coefficients $W^{s,t}_l$.\\
(1) Quasiparticle lifetime
\[
\tau(0)=\frac{4k^2_F\hbar^3}{\pi^3m^*k_B^2T^2}\frac{1}{\langle W(\theta,\phi)\rangle}, \nonumber\\
\]
(2) Thermal conductivity
\begin{eqnarray}
\kappa&=&(\pi^2/2)nk_B(T/T_F)v^2_F\tau(0)S_E(\lambda_\kappa), \nonumber\\
\lambda_\kappa&=&\langle W(\theta,\phi)(1+2\cos\theta)\rangle /\langle W(\theta,\phi)\rangle, \nonumber
\end{eqnarray}
(3) Viscosity
\begin{eqnarray}
\eta&=&\frac{1}{5}nv_Fp_F\tau(0)S_O(\lambda_\eta),\nonumber\\
\lambda_\eta&=&\langle W(\theta,\phi)(1-3\sin^4(\theta/2)\sin^2\phi)\rangle /\langle W(\theta,\phi)\rangle, \nonumber
\end{eqnarray}
(4) Spin diffusion
\begin{eqnarray}
D&=&\frac{1}{3}v^2_F(1+F^a_0)\tau(0)S_O(\lambda_D), \nonumber\\
1-\lambda_D&=&\langle W_{\uparrow\downarrow}(\theta,\phi)\sin^2(\theta/2)(1-\cos\phi)\rangle /\langle W(\theta,\phi)\rangle, \nonumber
\end{eqnarray}
The scattering rates, $W(\theta,\phi)$ and $W_{\uparrow\downarrow}(\theta,\phi)$, and the functions $S_{E(O)}(\lambda)$ are given by \cite{SS}
\begin{eqnarray}
W(\theta,\phi)&=&\frac{3}{8}T_t(\theta,\phi)^2+\frac{1}{8}T_s(\theta,\phi)^2+\frac{1}{4}T_t(\theta,\phi)T_s(\theta,\phi), \nonumber\\
W_{\uparrow\downarrow}(\theta,\phi)&=&\frac{1}{4}T_t(\theta,\phi)^2+\frac{1}{4}T_s(\theta,\phi)^2+\frac{1}{2}T_t(\theta,\phi)T_s(\theta,\phi), \nonumber\\
S_{E(O)}&=&\sum^{\infty}_{n={\rm even (odd)}}\frac{2n+1}{n(n+1)[n(n+1)-2\lambda]}. \nonumber
\end{eqnarray}
(5) Landau parameters $F_0^s$, $F_0^a$, and $F_1^s$
\begin{eqnarray}
\frac{F^s_l}{1+F^s_l/(2l+1)}&=&\frac{1}{4}\big\{[3W^t_l+W^s_l]\delta_{l,0}-(3W^t_l-W^s_l)\big\}, \nonumber\\
\frac{F^a_l}{1+F^a_l/(2l+1)}&=&\frac{1}{4}\big\{[W^t_l-W^s_l]\delta_{l,0}-(W^t_l+W^s_l)\big\}, \nonumber
\end{eqnarray}
where $W_{t,s}(1) = \sum^{\infty}_{l=0}W^{t,s}_l$.

We expand the scattering amplitude up to $l=3$,
\begin{eqnarray}
T_s(x_2,x_3)&=&2W^s_0+W^s_1\big(P_1(x_2)+P_1(x_3)\big)+W^s_2\big(P_2(x_2)+P_2(x_3)\big)+W^s_3\big(P_3(x_2)+P_3(x_3)\big)\nonumber\\
T_t(x_2,x_3)&=&W^t_1\big(P_1(x_2)-P_1(x_3)\big)+W^t_2\big(P_2(x_2)-P_2(x_3)\big)+W^t_3\big(P_3(x_2)-P_3(x_3)\big). 
\end{eqnarray}
Then, we have seven physical parameters, $\tau(0)T^2$, $\lambda_\kappa$, $\lambda_\eta$, $\lambda_D$, $F_0^s$, $F_0^a$, and $F_1^s$, and adjust the seven fitting parameters $W^{s,t}_l(l\leq 3)$ to minimize the sum of squared deviations of the calculated physical quantities from their corresponding experimental values. Using eq.(10), we can express $T^{\rm(s)}$ and $T^{\rm (a)}$ in terms of $T_t$ and $T_s$ as 
\begin{eqnarray}
T^{\rm (a)}(\theta,\phi)=\frac{1}{4}\big(T_t(x_2,x_3)-T_s(x_2,x_3)\big), &\quad& T^{\rm (s)}(\theta,\phi)=\frac{1}{4}\big(3T_t(x_2,x_3)+T_s(x_2,x_3)\big), \nonumber\\
T^{\rm (a)}(\theta',\phi')=\frac{1}{4}\big(T_t(x'_2,x'_3)-T_s(x'_2,x'_3)\big), &\quad& T^{\rm (s)}(\theta',\phi')=\frac{1}{4}\big(3T_t(x'_2,x'_3)+T_s(x'_2,x'_3)\big), 
\end{eqnarray}
and calculate the SC contribution $\delta\beta_j$ concretely. Including a frequency cutoff $\varepsilon_c$ in the gap function $\Delta$ in the form \cite{cutoff} $\Delta(\varepsilon_n;T)=\Delta(T)/[1+(\varepsilon_n/\varepsilon_c)^4]$, we obtain the following expression of the SC correction to $\beta_j$ in the relaxation time approximation 
\begin{equation}\label{eq.deltabetajrtaa}
\delta\beta_j = - \beta_0(T)\frac{T}{E_F}\Big[l_j^{\rm C}\tilde{S}_C+l_j^{\rm E+F}\tilde{S}_{E+F}+l_j^{\rm D}\tilde{S}_D\Big], 
\end{equation}
where the coefficients $l_j^{\rm K}$ are given in Tables V
. 
\begin{table}[h]
\begin{center}
\caption{Numerical values of the coefficients $l_j^{\rm K}$
.}
\begin{tabular}{c  c c  c | c c c | c c c}
\hline\hline
   & &P=12 (bar) & & & P=16 (bar) & & & P=20 (bar) & \\
\hline
j& $l_j^{\rm C}$    & $l_j^{\rm E+F}$ & $l_j^{\rm D}$ & $l_j^{\rm C}$ & $l_j^{\rm E+F}$ &$l_j^{\rm D}$ & $l_j^{\rm C}$  & $l_j^{\rm E+F}$ & $l_j^{\rm D}$ \\
\hline
\\
1 & 0.8543  & 0.6598 & 1.0695 & 1.0746 & 0.7405 & 1.2171 & 1.1026 & 0.7588 & 1.2137 \\
2 & -1.4707 & 15.3879 & -2.4577 & -1.6729 & 17.5695 & -2.8381 &-1.7178 & 18.1001 & -3.0316 \\
3 & 1.4660  & 3.1556 & 3.0504 & 1.7383 & 3.3996 & 3.5672 & 1.7339 & 3.4850 &3.5957 \\
4 & 3.9382 & 1.7999 & 6.8452 & 4.5471 & 1.9915 & 8.0498 & 4.6242 & 1.9306 & 8.2253 \\
5 & 7.7364  & 2.6753 & -1.6559 & 8.2614 & 2.8890 & -2.0485 &8.5912 & 3.0719 & -2.2023 \\
\\
\hline\hline
\\
\end{tabular}
\begin{tabular}{c  c c  c | c c c | c c c}
\\
\hline\hline
 & &P=24 (bar) & & & P=28 (bar) & & & P=34.4 (bar) & \\
\hline
j& $l_j^{\rm C}$    & $l_j^{\rm E+F}$ & $l_j^{\rm D}$ & $l_j^{\rm C}$ & $l_j^{\rm E+F}$ &$l_j^{\rm D}$ & $l_j^{\rm C}$  & $l_j^{\rm E+F}$ & $l_j^{\rm D}$ \\
\hline
\\
1 & 1.2522  & 0.9086 & 1.0230 & 1.3338 & 0.9154 & 0.9880 & 1.1269 & 0.9168 & 0.9559 \\
2 & -2.1025 & 19.9582 & -4.9303 & -2.1983 & 21.3248 & -5.9904 &-2.1791 & 20.0521 & -5.4358 \\
3 & 1.1921  & 3.3555 & 2.8961 & 0.9999 & 3.3166 & 2.5887 & 0.8187 & 3.3170 & 2.3066 \\
4 & 4.5375 & 0.5847 & 8.2523 & 4.4206 & -0.2511 & 8.6678 & 4.2128 & -0.1348 & 7.4739 \\
5 & 10.2905  & 4.5880 & -3.3103 & 11.1117 & 5.3985 & -4.1031 & 11.1826 & 5.2853 & -3.2555 \\
\\
\hline\hline
\end{tabular}
\end{center}
\end{table}   

The SC correction to $\beta_j$ in clean limit following from eq.(\ref{eq.deltabetajrtaa}) is given by 
\begin{equation}
\delta\beta_j = - \beta_0(T)\frac{T}{T_{c0}}\Delta{\beta_j}.
\end{equation}
Numerical values of $\Delta{\beta_j}$, in which the factor $T_{c0}/E_F$ is also included to collect the pressure dependence, are shown in Table.\ref{table:SS} where we set the cutoff $\varepsilon_c=0.068E_F$. These numerical values are almost same result as that in Ref.\cite{SS}. As one can see from the relative difference $|\Delta{\beta}_{\rm ABM}-\Delta{\beta}_{\rm BW}|$, PCP is obtained to be 24 bar which is close to the experimental PCP, 22 bar. 
\begin{table}[h]
\begin{center}
\caption{Numerical results of $\Delta \beta_j$ in eq.(31). \label{table:SS}}
\begin{tabular}{c c c c c c c c }
\hline\hline
\\
P[bar]& $\Delta{\beta_1}$    & $\Delta{\beta_2}$ & $\Delta{\beta_3}$ & $\Delta{\beta_4}$ & $\Delta{\beta_5}$ &$\Delta{\beta}_{\rm ABM}$ &$\Delta{\beta}_{\rm BW}$ \\
\\
\hline
\\
12 & 0.043  & 0.066 & 0.105 & 0.208 & 0.229 & 0.503 & 0.289 \\
16 & 0.057 & 0.087 & 0.135 & 0.267 & 0.267 & 0.621 & 0.367 \\
20 & 0.063  & 0.099 & 0.147 & 0.293 & 0.298 & 0.690 & 0.408 \\
24 & 0.071 & 0.088 & 0.125 & 0.291 & 0.376 & 0.755 & 0.423 \\
28 & 0.077  & 0.092 & 0.119 & 0.295 & 0.421 & 0.808 & 0.447 \\
34.4 & 0.073 & 0.096 & 0.115 & 0.288 & 0.461 & 0.845 & 0.457 \\
\\
\hline\hline
\end{tabular}
\end{center}
\end{table}   

In the present case, the new impurity-induced SC correction to $\beta_j$, eq.(\ref{eq:strongimp}), is given by 
\begin{eqnarray}
\delta{\tilde \beta}_j &=& - \frac{4\pi^2}{7\xi(3)}\beta_0(T)\Big(\sum_{n\geq 0}\frac{1}{\big(n+\frac{1}{2}+\frac{1}{4\pi\tau T}\big)^2}\frac{1}{[1+(\varepsilon_n/\varepsilon_c)^4]^2}\Big)^2\frac{1}{4\pi\tau T_{c0}}\Delta\tilde{\beta_j},
\end{eqnarray}
where $\Delta\tilde{\beta_j}$ are shown in Table.\ref{table:SSimp}. In comparison with Table.\ref{table:SS} all signs of $\Delta\tilde{\beta_j}$ are opposite to those of $\Delta{\beta_j}$, implying that the impurity scattering weakens the SC corrections for both the ABM and BW state. As for the relative stability of the pairing states, the impurity scattering favors the BW pairing state since the relation $|\Delta\tilde{\beta}_{\rm ABM}|>|\Delta\tilde{\beta}_{\rm BW}|$ is always satisfied.

\begin{table}[h]
\begin{center}
\caption{Numerical results on $\Delta {\tilde \beta}_j$ in eq.(32). \label{table:SSimp}}
\begin{tabular}{c c c c c c c c }
\hline\hline
\\
P[bar]& $\Delta\tilde{\beta_1}[10^{-3}]$    & $\Delta\tilde{\beta_2}[10^{-3}]$ & $\Delta\tilde{\beta_3}[10^{-3}]$ & $\Delta\tilde{\beta_4}[10^{-3}]$ & $\Delta\tilde{\beta_5}[10^{-3}]$ &$\Delta\tilde{\beta}_{\rm ABM}[10^{-3}]$ &$\Delta\tilde{\beta}_{\rm BW}[10^{-3}]$ \\
\\
\hline
\\
12 & -0.378  & -2.360 & -2.047 & -3.292 & -1.374 & -7.026 & -4.975 \\
16 & -0.642 & -3.004 & -2.559 & -4.033 & -1.868 & -8.904 & -6.466 \\
20 & -0.719  & -3.355 & -2.862 & -4.538 & -2.131 & -10.024 & -7.252 \\
24 & -0.821 & -3.106 & -2.643 & -4.563 & -2.637 & -10.306 & -7.209 \\
28 & -0.957  & -3.074 & -2.623 & -4.670 & -3.059 & -10.804 & -7.482 \\
34.4 & -0.688 & -3.069 & -2.670 & -4.999 & -2.907 & -10.975 & -7.282 \\
\\
\hline\hline
\end{tabular}
\end{center}
\end{table}

\begin{figure}[h]
\includegraphics[scale=0.27]{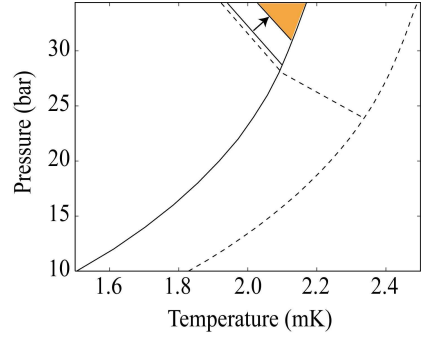}
\caption{The corresponding result to Fig.6 in the phenomenological approach obtained in terms of $1/(2\pi\tau)=0.13$ (mK). In comparison with the case of the perturbative approach, the shrinkage of the ABM pairing region is much more remarkable. \label{fig:SS}}
\end{figure}

The {\it mean field} P-T phase diagram in the case with no anisotropic scattering effect is shown in Fig.\ref{fig:SS} where the solid lines are $T_{AB}(P)$ in the impure case and result from the use of the value $1/(2\pi\tau)=0.13$ (mK), while the dashed line is that for the bulk liquid (i.e., in clean limit). In obtaining the curves in this figure, we have used the same experimental data as in the previous work \cite{SS}. Then, a necessary set of Landau parameters was available only at three pressure values, 24, 28, and 34.4 (bar) above the bulk polycritical point (PCP). This is why the resulting bulk $T_{AB}(P)$ curve has an inessential kink in the figure. 
The lower (upper) solid line, implying an AB transition curve, is obtained without (with) the impurity-induced SC correction given by eq.(18). As one can see in Fig.\ref{fig:SS}, the ABM region shrinks as the impurity scattering is enhanced. It is important to note that the $T_{\rm AB}$ line obtained with $\delta{\tilde \beta}_j$ lies at much higher pressures than the other corresponding curves obtained in clean limit, in the spin-fluctuation approach, and in the approach used in the previous subsection. It suggests that the use of this phenomenological approach explains better the experimentally determined global phase diagrams of liquid $^3$He not only in clean limit \cite{SS} but also in the weakly disordered case, i.e., in aerogel.

\section{Application to uniaxially anisotropic aerogel}

Since the coherence length $\xi_0$ of superfluid $^3$He at $T=0$ becomes shorter as the pressure increases, $\xi_0$ can be much shorter than the structural correlation length of aerogel $\xi_a\approx 30-100$nm at high pressures (see sec.I). In this high pressure region, the orientation of scatterers is not random over the scale $\sim \xi_a$, and the Cooper pairs can be regarded as behaving in scattering potentials with a {\it fixed} uniaxial anisotropy. When this local anisotropy induced by the aerogel structure is well defined, the role \cite{AI} of aerogel acting as a "random pinning" of the order parameter field (such as the $l$-vector in the ABM state) is a minor correction to the free energy, and a mean field analysis assuming a fixed anisotropy in the scattering events is expected to well describe true pairing states of the A-like and B-like phases in aerogel which is locally anisotropic but globally isotropic. \cite{AI2} Or, it is possible that an aerogel sample used in experiments has accidentally or artificially a global anisotropy. Commonly to these two situations, the mean field analysis assuming a small but nonvanishing anisotropy in the impurity-scattering amplitude becomes an appropriate description \cite{AI2}. Based on these facts, we apply the expressions of the impurity induced SC corrections to the case of an uniaxially anisotropic aerogel by introducing \cite{AI2} an anisotropy in the scattering amplitude. In performing this, the effect of the anisotropy on the SC correction is neglected, because both of them are small effects, and we have only to incorporate $\delta\tilde{\beta_j}$ and $\delta{\beta_j}$ in the $\beta_j$ parameters in the analysis performed in Ref.\cite{AI2}. Thus, only the obtained results will be discussed in the remainder of this section.

In the previous paper \cite{AI2} we found that, in the T-P phase diagram, the ABM pairing region becomes wider in both uniaxially stretched and compressed aerogels and, especially, that the polar pairing state appears near $T_c(P)$ as a 3D phase in the uniaxially stretched case. As is easily expected, the stability of the ABM pairing state is weakened by including the impurity induced SC correction $\delta\tilde{\beta_j}$ and, as a result, the ABM pairing region will shrink. In Fig.8, a result of the uniaxially stretched case based on the perturbative approach is shown, where the parameter values $(2\pi\tau)^{-1}=0.13$ (mK) and $\lambda=0.821$ are used together with the anisotropy parameter \cite{AI2} $\delta_u=-0.06$. Since the polar pairing state is induced by the $quadratic$ term of the GL free energy functional, its region is unaffected by the inclusion of $\delta\tilde{\beta_j}$ in contrast to the shrinkage, indicated by an arrow, of the ABM pairing region at high pressures. A similar feature can also be seen in the uniaxially compressed case. Further, although we have shown here only the result in the perturbative approach, we have verified that essentially the same result as above is obtained in the phenomenological approach.

\begin{figure}[h]
\includegraphics[scale=0.32]{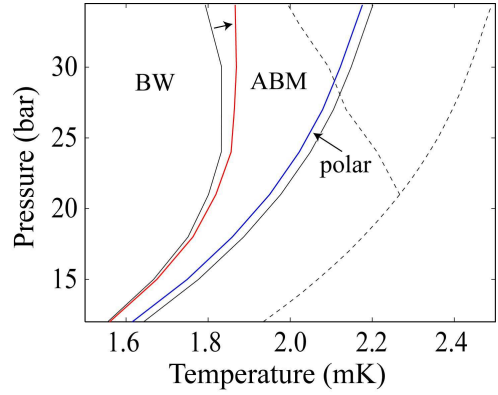}
\caption{Phase diagram in the perturbative approach for the uniaxially stretched case obtained in terms of $(2\pi\tau)^{-1}=0.13$ (mK), $\lambda=0.821$, and $\delta_u=-0.06$. Since the polar pairing state is determined by the quadratic term in GL free energy functional, its region is unaffected by changing the SC correction in $\beta_j$. In contrast, the ABM pairing region shrinks due to the inclusion of $\delta\tilde{\beta_j}$.}
\end{figure}
Next we investigate the case with a {\it weaker} anisotropy in order to obtain a comparable result with $^3$He in a realistic aerogel. We choose the anisotropy value $\delta_u=-0.005$ and compare the result in the perturbative approach and that in the phenomenological approach with each other. Those results are shown in Fig.9. Commonly in both the approaches, the magnitude of the anisotropy $|\delta_u|$ is so small that the polar pairing region is too narrow to become visible in the T-P phase diagram. The ABM pairing region survives even at lower pressures than the mean field "triangle" (see Figs.3 and 7), and the PCP doesn't exist. At high pressures, the slope $dT_{\rm AB}/dP$ becomes not negative but positive. It suggests that the previous result, such as Fig.1 in Ref.\cite{AI}, at higher pressures will be improved if $\delta\tilde{\beta_j}$ is included there. An important difference between the two approaches is also seen in $T_{\rm AB}(P)$ at the highest pressure, $P=34.4$ bar, in the figures. Reflecting the contribution of $\delta\tilde{\beta_j}$, $T_{\rm AB}$ at 34.4 (bar) in the right figure of Fig.9 following from the phenomenological approach is higher than that of the bulk liquid. It implies that the features of the T-P phase diagram obtained in the phenomenological approach \cite{SS} are in better agreement with experimental ones than those in other approaches.

Furthermore, the shaded region of the right diagram in Fig.9 resembles the experimentally obtained region of the A-like phase in Ref.4 where $d T_{\rm AB}/dP$ is positive even at much higher pressures than PCP estimated there \cite{Osheroff}. It might imply that a global anisotropy in aerogel samples, created through their production processes, needs to be taken into account in understanding the experimental phase diagram. That is, it is possible that the window of the A-like phase at lower pressures is too narrow to be observed experimentally. Alternatively, at lower pressures where $\xi_0 \gg \xi_a$ may be satisfied, a $T_c$ shift due to the quenched disorder \cite{AI} outweighs that due to the anisotropy, and the scenario in Ref.9 invoking the A-like phase induced by the quenched disorder may be more appropriate. 
\begin{figure}[t]
\includegraphics[scale=0.55]{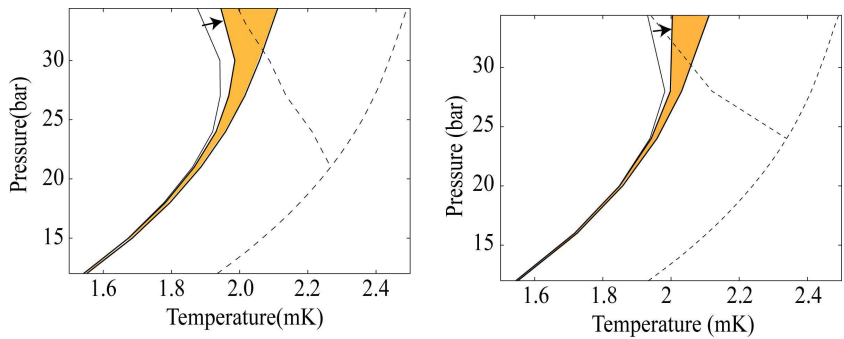}
\caption{Phase diagrams in the uniaxially stretched case obtained in the perturbative approach (left) for $(2\pi\tau)^{-1}=0.15$ (mK), $\lambda=0.821$, and $\delta_u=-0.005$ and in the phenomenological approach (right) for $(2\pi\tau)^{-1}=0.15$ (mK) and $\delta_u=-0.005$. Each arrow indicates a shift of $T_{\rm AB}(P)$ due to the inclusion of $\delta\tilde{\beta_j}$. Due to its inclusion, the slope of $T_{\rm AB}(P)$ can easily become positive at least in the phenomenological approach. Although the slope in the left figure is not positive, a positive slope can be achieved by increasing $\tau^{-1}$ value for the fixed $\delta_u$ even in the perturbative approach.}
\end{figure}

\section{Conclusion}

We have investigated the impurity-induced strong coupling correction up to the lowest order in $1/\tau T$ by taking account of additional contributions overlooked in the relaxation time approximation and have calculated such contributions to the quartic term of GL free energy functional using the perturbative approach in a short range repulsive interaction and the phenomenological approach. In both approaches, the impurity scattering favors the BW pairing state, and the ABM region shrinks just as suggested experimentally \cite{Halperin,A1A2}. The resulting extent of the shrinkage in the phenomenological approach \cite{SS} is especially substantial and seems to be consistent with the expectation \cite{Halperin}. 

Here, several approximations used in our calculations will be discussed. First, our derivation of the AB transition line based on the GL free energy with no sixth order term is not conventional. Although this treatment might have a subtle aspect in obtaining the bulk AB transition curve far apart from $T_c(P)$, it is safely valid for the present purpose of obtaining the AB transition curve of the liquid $^3$He in aerogel occurring near $T_c$. 
In our treatment on the quasiparticle scatterings, angular dependences of the scattering amplitude were neglected. For the scattering events in the locally anisotropic aerogels, this approximation of the $s$-wave scattering may be too simple, and we cannot necessarily exclude a possibility that it might have led to different angular dependences significantly affecting the impurity-induced SC correction $\delta\tilde{\beta_j}$. Further, only the lowest order terms in the impurity scattering were kept in our calculation by assuming $T \tau > 1$. Although this treatment is safely valid for the present purpose focusing primarily on the high pressure region in which the SC effect is more important, in low enough pressures along the $T_c(P)$-curve or close to the quantum critical point, the above inequality is not satisfied, and rather, calculations need to be performed in the dirty limit. 

After performing microscopic calculations in the case with an isotropic scattering amplitude according to the two approaches, the present theory has been applied to superfluid $^3$He in aerogel by assuming the aerogel sample to have an uniaxial {\it anisotropy} in scattering events over length scales longer than the coherence length $\xi_0$. We find that the unexpected positive slope of the AB transition curve \cite{Osheroff} can be obtained at high pressures, and that the ABM pairing region survives even at low pressures with a narrow width. The assumption of an uniaxial anisotropy over large scales is realistic from two different points of view. First, as mentioned in sec.I, the relation $\xi_0 \ll \xi_a$ may be satisfied in a globally {\it isotropic} aerogel under higher pressures, and then, the condensation energy is well approximated \cite{AI2} by that in a globally {\it anisotropic} case. Second, aerogel samples can naturally or artificially \cite{Halperin2,Matsubara} have a global anisotropy as a result of their production process. For either of such two situations, the phenomenological approach \cite{SS} gives more consistent results with experimental facts compared with other models. Consequently, we feel that an identification \cite{AI,Volovik} between a superfluid glass phase with the ABM pairing and the A-like phase has obtained an additional support.

\begin{acknowledgements}
We are grateful to Y. Lee for discussions, and one of authors (R.I.) thanks J. A. Sauls for a discussion on the SC correction on which our analysis using the approach in Ref.\cite{SS} is based. The numerical calculations were carried out at YIFP in Kyoto University. 
This work is financially supported by a Grant-in-Aid from the Ministry of Education, Culture, Sports, Science, and Technology, Japan. 
\end{acknowledgements}


\begin{thebibliography}{99}
\bibitem{com1} If the A-like phase occurs as a nonequilibrium supercooled state, $T_{\rm AB}(P)$ line seems to still show a negative slope.
\end{thebibliography}
\end{document}